\documentclass[aps, prd, superscriptaddress, preprintnumbers, floatfix,10pt]{revtex4-2}

\usepackage{graphicx,amsfonts,amsmath,amssymb,amstext}
\usepackage{float,wrapfig}
\usepackage{subfigure, psfrag}
\usepackage{dsfont}
\usepackage{color}
\usepackage[colorlinks=true,urlcolor=blue,citecolor=blue,linkcolor=blue]{hyperref}
\usepackage{bm}

\newcommand{\be}{\begin{equation}}
\newcommand{\ee}{\end{equation}}
\newcommand{\ba}{\begin{eqnarray}}
\newcommand{\ea}{\end{eqnarray}}
\newcommand{\sign}[1]{\,\mbox{sgn}\left({#1}\right)}

\definecolor{purple}{rgb}{0.8,0,0.6}
\definecolor{darkgreen}{rgb}{0.00,0.6,0.00}

\extrafloats{100}

\begin{document}

\title{Stray magnetic field and stability of time-dependent viscous electron flow}
\date{November 11, 2021}

\author{P.~O.~Sukhachov}
\email{pavlo.sukhachov@yale.edu}
\affiliation{Department of Physics, Yale University, New Haven, Connecticut 06520, USA}

\author{E.~V.~Gorbar}
\affiliation{Department of Physics, Taras Shevchenko National University of Kyiv, Kyiv, 03022, Ukraine}
\affiliation{Bogolyubov Institute for Theoretical Physics, Kyiv, 03143, Ukraine}

\begin{abstract}
A hydrodynamic flow of electrons driven by an oscillating electric field in Dirac and Weyl semimetals is investigated. It is found that a double-peak profile of the electric current appears and is manifested in a stray magnetic field with peaks in one of the field components. The nontrivial current profile originates from the interplay of viscous and inertial properties of the electron fluid as well as the boundary conditions. Analytical results are supported by numerical calculations in samples of different geometries such as straight channels, nozzles, and cavities. The double-peak profile of the current is found to be qualitatively insensitive to a specific form of the time dependence of the oscillating electric field and is stable with respect to the sample geometry. A phase diagram and criteria for observing the double-peak structure are determined. In addition, it is shown that nozzles and cavities provide an efficient means to locally enhance or reduce the fluid velocity.
\end{abstract}

\maketitle

\section{Introduction}
\label{sec:Introduction}

Transport properties are among the most fundamental characteristics of any material. Among several transport regimes, the hydrodynamic one draws significant attention. It is realized when the electron-electron interactions dominate over the scatterings of electrons on impurities and phonons. The hydrodynamic regime of charge and heat transport in solids was proposed in the 1960s~\cite{Gurzhi:1963,Gurzhi:1968,Nielsen-Shklovskii:1969,Nielsen-Shklovskii:1969a}. However, the first experimental signatures of hydrodynamic transport were observed only three decades later in the late 1990s in a two-dimensional (2D) electron gas of high-mobility $\mathrm{(Al,Ga)As}$ heterostructures~\cite{Molenkamp-Jong:1994,Jong-Molenkamp:1995}. In particular, it was shown that the resistivity decreases with temperature~\cite{Molenkamp-Jong:1994,Jong-Molenkamp:1995}, which is known as the Gurzhi effect~\cite{Gurzhi:1963}; see Ref.~\cite{Kashuba-Molenkamp:2018} for the Gurzhi effect in systems with relativisticlike dispersion relation. Later, the characteristic dependence of the resistivity on the channel size was observed in ultrapure 2D metal palladium cobaltate ($\mathrm{PdCoO_2}$)~\cite{Moll-Mackenzie:2016}, which agrees with the dependence expected for the Poiseuille flow of the electron fluid.

The realization of the hydrodynamic regime in graphene~\cite{Crossno-Fong:2016,Ghahari-Kim:2016,Krishna-Falkovich:2017,Berdyugin-Bandurin:2018,Bandurin-Falkovich:2018,Ku-Walsworth:2019,Sulpizio-Ilani:2019} had a strong impact on the development of electron hydrodynamics in solids~\cite{Lucas-Fong:rev-2017,Narozhny:rev-2019}; see also Ref.~\cite{Sante-Wehling:2020} for the recent prediction of the realization of electron hydrodynamics in kagome metals. As an example of an interesting effect related to a hydrodynamic flow, graphene's constriction can have higher conduction in the hydrodynamic regime than in the ballistic one~\cite{Krishna-Falkovich:2017,Guo-Levitov:2017}. Three-dimensional (3D) Dirac and Weyl semimetals, whose quasiparticles are described by the Dirac and Weyl equations, respectively, can be considered as 3D analogs of graphene~\cite{Armitage:rev-2018,GMSS:book}. The experimental observation of the dependence of the electric resistivity on the channel width and the violation of the Wiedemann--Franz law with the lowest Lorenz number ever reported indicate the realization of the hydrodynamic transport regime in the Weyl semimetal WP$_2$~\cite{Gooth-Felser:2018}. Recently, a hydrodynamic profile of the direct electric current was visualized via stray magnetic fields in the Weyl semimetal WTe$_2$~\cite{Vool-Yacoby:2020-WTe2}.

In the majority of experimental and theoretical studies of hydrodynamic transport, the current is driven by a static voltage. For example, the steady Poiseuille flow in channels was studied in numerous works including, e.g., Refs.~\cite{Jong-Molenkamp:1995,Torre-Polini:2015,Gorbar:2018vuh,Kashuba-Molenkamp:2018,Erdmenger-Fernandez:2018,Gooth-Felser:2018,Ku-Walsworth:2019,Sulpizio-Ilani:2019,Vool-Yacoby:2020-WTe2}.
A time-dependent or pulsating flow was investigated in Refs.~\cite{Moessner-Witkowski:2018,Moessner-Witkowski:2019}. It was shown that the maximum of the flow velocity migrates from the center toward the edges where boundary layers are formed. As an experimentally accessible signature of these layers, a nontrivial dependence of the optical conductivity on the sample size and frequency was proposed.

Motivated by significant attention to the dynamical properties of materials, we study the response of electron fluid in the hydrodynamic regime in both 3D and 2D materials with a relativisticlike dispersion relation to a time-dependent drive and identify the corresponding signatures in stray magnetic fields. We confirm that a nontrivial double-peak hydrodynamic profile of the electric current appears in the transient regime. These peaks originate from the interplay of the inertial and viscous properties of the electron fluid. We identify the phase diagram, i.e., the parameter range where one expects the appearance of the double-peak structure. Moreover, the stability of the double-peak structure with respect to the boundary conditions, the channel geometry, and the time profile of the drive is investigated. While the profiles of the electric current are distorted in nozzles and cavities, the double-peak structure remains as long as the fluid sticks to the surfaces of the channel (no-slip boundary conditions) and the driving force changes its sign.

We propose to probe the nontrivial current distribution via stray magnetic fields. It is shown that the double-peak current profile leads to a nonmonotonic magnetic field distribution where the in-plane component of the field perpendicular to the flow also acquires peaks near the boundaries. Our estimates suggest that the magnitude of the magnetic fields can be within the reach of modern magnetometry techniques such as the quantum spin magnetometry (QSM)~\cite{Maze-Lukin:2008-QSM,Levine-Walsworth:2019-QSM} and the scanning superconducting quantum interference device (SQUID)~\cite{Cui-Moler:2017} techniques. Thus, the key findings of this paper are (i) a nontrivial distribution of the magnetic field caused by the double-peak electric current profile in the transient regime; (ii) the stability of the double-peak current profile, in particular, with respect to the geometry of the channel; and (iii) the identification of the phase diagram and the criteria for observing the double-peak profile.

The paper is organized as follows. The model, main equations, and boundary conditions are described in Sec.~\ref{sec:model}. Steady flows are studied by using both analytical and numerical approaches in Sec.~\ref{sec:steady}. Section~\ref{sec:dynamic} is devoted to electron fluid flows driven by an oscillating driving force. Stray magnetic fields generated by pulsating viscous electron flows are considered in Sec.~\ref{sec:magnetic-field}. The obtained results are summarized in Sec.~\ref{sec:Summary}. Through this study, we set $k_{B}=1$.

\section{Model}
\label{sec:model}

In this section, we present the key equations, define the boundary conditions, and discuss the model setup.

\subsection{Hydrodynamic equations and model setup}
\label{sec:model-eqs}

The hydrodynamic equations for the electron fluid include the Navier-Stokes, energy continuity, and electric charge conservation equations. In addition, since the electron fluid is the charged one, one should include Maxwell's equations. Their role, however, is less profound in transport than in, e.g., collective modes~\cite{Lucas-Fong:rev-2017} or convection~\cite{Sukhachov-Shovkovy:2021}. In this paper, we consider only slow flows, where the electron fluid velocity $\mathbf{u}$ is much smaller than the Fermi velocity $v_F$. Therefore, we linearize the hydrodynamic equations for the relativisticlike electron fluid in Dirac and Weyl semimetals and, in the quasistatic approximation, retain only the Gauss law. The Ampere law does not affect hydrodynamic flows in a linearized regime and will be considered only in Sec.~\ref{sec:magnetic-field} where stray magnetic fields are calculated. The resulting system reads as~\cite{Lucas-Fong:rev-2017,Narozhny:rev-2019}
\begin{eqnarray}
\label{model-Eq1}
&&\frac{w_0}{v_F^2}\partial_t\mathbf{u} -\eta \left(\Delta -\frac{w_0}{v_F^2\tau \eta}\right)\mathbf{u} -\frac{\eta (d-2)}{d} \bm{\nabla}\left(\bm{\nabla}\cdot\mathbf{u}\right) +\bm{\nabla}P= -en_0\mathbf{E},\\
\label{model-Eq2}
&&\partial_t\epsilon +w_0(\bm{\nabla}\cdot\mathbf{u}) =0,\\
\label{model-Eq3}
&&-e\partial_tn -en_0\left(\bm{\nabla}\cdot\mathbf{u}\right) +\sigma \left(-\Delta \varphi +\frac{1}{e} \Delta \mu -\frac{\mu_0}{eT_0} \Delta T\right)=0,\\
\label{model-Eq4}
&&\left(\bm{\nabla} \cdot \mathbf{E}\right) = -4\pi e (n-n_0).
\end{eqnarray}
Here, $w=\epsilon+P$ is the enthalpy, $\epsilon$ is the energy density, $P$ is the pressure, $n$ is the electron number density, $\mathbf{E}$ is the electric field, $v_F$ is the Fermi velocity, $\mu$ is the chemical potential, $T$ is temperature, $\tau$ is the relaxation time describing the relaxation of the total momentum of the fluid~\cite{Gurzhi:1968}, $\varphi$ is the electric potential, and $-e$ is the electron charge. The shear viscosity is denoted as $\eta$ and we omit the bulk viscosity, which is negligible in relativisticlike systems (see, e.g., Ref.~\cite{Principi-Polini:2015}). One has the following relation for the shear viscosity: $\eta = \eta_{\rm kin}w_0/v_F^2$, where $\eta_{\rm kin}\sim v_F^2\tau_{\rm ee}$ is the kinematic shear viscosity and $\tau_{\rm ee}$ is the electron-electron scattering time. In graphene, $\eta_{\rm kin}= v_F^2\tau_{\rm ee}/4$~\cite{Alekseev:2016}. For a relativisticlike fluid, $P=\epsilon/d$ and $w=(d+1)\epsilon/d$ where $d=2,3$ is the spatial dimension. In our numerical estimates, we use $\tau_{\rm ee}\approx \hbar/T$ reported for WP$_2$~\cite{Gooth-Felser:2018}. Finally, subscript $0$ denotes the global equilibrium values of parameters. The expressions for $n_0$ and $\epsilon_0$ are given in Appendix~\ref{sec:app-1}.

The linearized electric and energy currents in the hydrodynamic regime are given by (see, e.g., Refs.~\cite{Lucas-Fong:rev-2017,Erdmenger-Fernandez:2018})
\begin{equation}
\label{model-J-lin}
\mathbf{J} = -en_0\mathbf{u} +\sigma \left(\mathbf{E} +\frac{1}{e}\bm{\nabla}\mu -\frac{\mu_0}{eT_0}\bm{\nabla}T\right) \quad \mbox{and} \quad \mathbf{J}^{\epsilon} = w_0\mathbf{u},
\end{equation}
respectively.

The intrinsic conductivity $\sigma$ characterizes the quantum critical contribution to the electric current. It can be estimated in a holographic approach~\cite{Hartnoll-Sachdev:2007,Kovtun-Ritz:2008,Hartnoll:2014,Landsteiner-Sun:2015,Davison-Hartnoll:2015} and reads as
\begin{eqnarray}
\label{model-sigma-3D}
&&\mbox{3D:} \quad \sigma = \frac{3 \pi\hbar v_F^3}{2} (\partial_{\mu} n) \tau_{\rm ee},\\
&&\mbox{2D:} \quad \sigma = \frac{2e^2}{\pi \hbar}.
\end{eqnarray}

Let us now discuss the model setup. We consider three geometries: (i) straight channel, (ii) nozzle, and (iii) cavity. Nozzle and cavity are schematically shown in Figs.~\ref{fig:steady-num-u-2D}(b) and \ref{fig:steady-num-u-2D}(c), respectively. In our numerical calculations, we assume that samples have length $L_x$ along the $x$-direction and their width is $L_y$ at $x=0$ and $x=L_x$. The shape of the boundaries is defined analytically by the following expressions:
\begin{equation}
\label{model-nozzle-shape}
y= \frac{L_N}{2} \left[e^{-\left(x-L_x/2\right)^2} - e^{-L_x^2/4}\right] \quad \mbox{and} \quad y=L_y-\frac{L_N}{2} \left[e^{-\left(x-L_x/2\right)^2} - e^{-L_x^2/4}\right],
\end{equation}
for the nozzle ($L_{N}>0$) and cavity ($L_{N}<0$) constrictions. Clearly, the width in the middle of the long channels ($x=L_x/2$) is $L_y-L_{N}$. The case of a straight channel corresponds to $L_{N}=0$. For simplicity, we assume that all samples are infinite along the $z$ direction in 3D. The same geometries are considered for 2D materials where the $z$ direction is omitted.

\subsection{Driving force, boundary conditions, and numerical parameters}
\label{sec:model-eqs-BC}

\begin{figure*}[ht]
\centering
\includegraphics[width=0.35\textwidth]{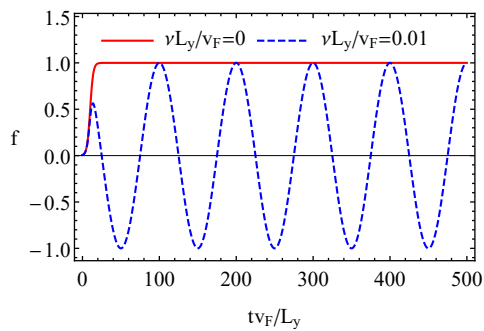}
\caption{The dependence of $f(t)$ on time $t$ given by Eq.~(\ref{model-f-BC}) at $\nu=0$ (red solid line) and $\nu=0.01\, v_F/L_y\approx22~\mbox{MHz}$ (blue dashed line). Other parameters are defined in Eqs.~(\ref{model-num-par-1}) and (\ref{model-num-par-2}).
}
\label{fig:model-f}
\end{figure*}

An important ingredient needed for the description of hydrodynamic flows in finite samples is boundary conditions. We consider a time-dependent driving force via the voltage applied to the side surfaces of the sample,
\begin{equation}
\label{model-phi-BC}
\varphi(t,x=0,y,z) = 0, \quad \varphi(t,x=L_x,y,z) = -E_0 f(t) L_x,
\end{equation}
where $E_0$ is the magnitude of the electric field and
\begin{equation}
\label{model-f-BC}
f(t) = \frac{1}{1+e^{\xi (t_{o}-t)}} \cos{(2\pi \nu t)}
\end{equation}
defines the time dependence of the applied voltage (a wavelet profile of the driving force is considered in Appendix~\ref{sec:app-3}). Here, $\nu$ is the frequency of oscillations, $t_o$ corresponds to the offset
time, and $\xi$ defines the steepness of the initial increase. In the analytical analysis, we set $\xi\to\infty$ and $t_{o}\to0$. The function $f(t)$ is plotted in Fig.~\ref{fig:model-f}.

In a general case, temperature at the ends of the channel can be different,
\begin{equation}
\label{model-T-BC}
T(t,x=0,y,z) = T_{\rm L}, \quad \quad  T(t,x=L_x,y,z) = T_{\rm R}.
\end{equation}
As to the electron density, we assume also that it is fixed at the boundaries in 3D samples
\begin{equation}
\label{model-n-BC}
n(t,x=0,y,z) = n(t,x=L_x,y,z) = n_0.
\end{equation}

In the 2D case, to determine the electric field in the sample, we employ the ``gradual channel" approximation~\cite{Shur:book,Dyakonov-Shur:1993}. In this case, instead of finding a solution of the Gauss law (\ref{model-Eq4}), the electric field is directly related to the electron density in the sample,
\begin{equation}
\label{model-graphene-capacitance}
\mathbf{E}= \frac{e}{C} \bm{\nabla} n,
\end{equation}
where $C=\varepsilon/(4\pi L_g)$ is the capacitance per unit area, $\varepsilon$ is the dielectric constant of the substrate, and $L_g$ is the distance to the gate. Therefore, we have
\begin{equation}
\label{model-graphene-n-BC}
n(t,x=0,y) = n_0, \quad  \quad n(t,x=L_x,y) = n_0 +\frac{C}{e} E_0 f(t) L_x
\end{equation}
in the 2D case.

As for the fluid velocity, we assume that both normal and tangential components of the fluid velocity vanish at the surfaces or edges of the sample (the latter conditions are known as the no-slip boundary conditions~\cite{Landau:t6-2013}; see also Refs.~\cite{Kiselev-Schmalian:2019,Moessner-Witkowski:2019} for detailed discussion in the context of the electron hydrodynamics in solids),
\begin{equation}
\label{model-ux-BC}
\left(\hat{\mathbf{n}}\cdot\mathbf{u}\right)\Big|_{\left\{x,y\right\} \in \mathcal{S}}=0 \quad \mbox{and} \quad
\left[\hat{\mathbf{n}}\times\mathbf{u}\right]\Big|_{\left\{x,y\right\} \in \mathcal{S}}=\mathbf{0},
\end{equation}
where $\hat{\mathbf{n}}$ is the surface normal and $\mathcal{S}$ denotes the surface of slab in 3D or edges in a 2D ribbon. The case of more general boundary conditions is briefly addressed in Appendix~\ref{sec:app-2}. Finally, the electric current should not flow through the sides of the channel, i.e.,
\begin{equation}
\label{model-Jx-BC}
\left(\hat{\mathbf{n}}\cdot\mathbf{J}\right)\Big|_{\left\{x,y\right\} \in \mathcal{S}}=0.
\end{equation}

Equations~(\ref{model-Eq1})--(\ref{model-J-lin}) together with the boundary conditions (\ref{model-phi-BC}), (\ref{model-T-BC}), (\ref{model-n-BC}) or (\ref{model-graphene-n-BC}), (\ref{model-ux-BC}), and (\ref{model-Jx-BC}) comprise a
system of equations that describes the hydrodynamic flow of a charged fluid in a channel.

In our numerical calculations of 3D hydrodynamic flows, we use material parameters for WP$_2$~\cite{Kumar-Felser:2017} and the following typical values of sample sizes and external fields:
\begin{eqnarray}
\label{model-num-par-1}
&&v_F=1.4\times10^7~\mbox{cm/s}, \quad \tau=0.3~\mbox{ns}, \quad T_0=10~\mbox{K}, \quad \mu_0 = 20~\mbox{meV}, \quad E_0=2~\mbox{mV/m},\\
\label{model-num-par-2}
&&L_y = 10~\mu\mbox{m}, \quad L_x=5\, L_y, \quad \xi =0.2\, v_F/L_y \approx 7~\mbox{ns}^{-1}, \quad t_{o}=10\, L_y/v_F \approx 0.71~\mbox{ns}.
\end{eqnarray}

In the 2D case, we use the parameters of graphene with $v_F=1.1\times10^{8}~\mbox{cm/s}$, $\mu_0=100~\mbox{meV}$, $T_0=100~\mbox{K}$, $\tau\approx0.1~\mbox{ns}$, and $L_{g}=100~\mbox{nm}$. In addition, we assume a hexagonal boron nitride substrate with the dielectric constant $\varepsilon\approx 3.3$.

As we show for a steady flow in Sec.~\ref{sec:steady}, temperature gradient could play a role similar to the electric field in driving the electron fluid flow. The effect of temperature gradient is, however, weak and leads to a small asymmetry of the electron fluid velocity with respect to $E_0 \to -E_0$. Therefore, to simplify the presentation, we take into account temperature gradient only in general expressions in Sec.~\ref{sec:steady}.

The case of a straight channel, which is infinite along the $x$ direction, can be easily analyzed analytically. This is done in Secs.~\ref{sec:steady} and \ref{sec:dynamic-analyt} for steady and time-dependent flows, respectively. In general, nozzle and cavity geometries admit only numerical solutions. The corresponding solutions are presented in Secs.~\ref{sec:steady} and \ref{sec:dynamic-num} for steady and dynamic flows, respectively.

\section{Steady flow}
\label{sec:steady}

As a warm-up, let us begin the analysis with steady flows for a static voltage, i.e., we set $\nu=0$ in Eq.~(\ref{model-f-BC}). The results obtained in this section allow us to better understand the role of the time-dependent driving force considered in Sec.~\ref{sec:dynamic}.

We start with the case of a steady flow in a channel infinite in the $x$ direction. In this case, if the ends of the channel are kept at different temperatures, then temperature gradient develops. The corresponding equation can be obtained by calculating the divergence of Eq.~(\ref{model-Eq1}) and using Eq.~(\ref{model-Eq3}). Notice that, in view of Eq.~(\ref{model-Eq2}), the electron fluid is incompressible. Then, after straightforward manipulations, we obtain
\begin{equation}
\label{steady-analyt-T-eq}
\Delta T(\mathbf{r})=0.
\end{equation}
The solution of this equation with the boundary conditions given in Eq.~(\ref{model-T-BC}) is simple
\begin{equation}
\label{steady-analyt-T1-sol}
T(\mathbf{r}) = T_{\rm L} +\frac{T_{\rm R}-T_{\rm L}}{L_x}x
\end{equation}
and corresponds to a constant temperature gradient $\partial_{x} T(\mathbf{r}) = \left(T_{\rm R}-T_{\rm L}\right)/L_x$ along the $x$ direction.

In infinite channels, the electric charge deviations from a background charge $n_0$ are suppressed due to the Gauss law, $n\approx0$~\footnote{Any deviations from local neutrality, where the neutrality is guaranteed by the compensating charge of immovable lattice ions, induce a strong electric field and, consequently, are energetically unfavorable.}. Therefore, in the linearized regime with $n = (\partial_{\mu} n) \mu +(\partial_{T} n) T$, we obtain $\mu \approx -T(\partial_{T}n)/(\partial_{\mu}n)$. Then, the following solution for the $x$ component of the fluid velocity follows from the $x$ component of Eq.~(\ref{model-Eq1}):
\begin{equation}
\label{steady-analyt-ux-sol}
u_{x}(y) = -\frac{\tau v_F^2 n_0}{w_0} \left\{eE_0 +\frac{1}{n_0} \left[s_0 -n_0 \frac{(\partial_{T}n)}{(\partial_{\mu}n)}\right] (\partial_{x} T)\right\} \left[1 - \frac{\cosh{\left(\frac{L_y-2y}{2\lambda_{G}}\right)}}{\cosh{\left(\frac{L_y}{2\lambda_{G}}\right)}}\right],
\end{equation}
where
\begin{equation}
\label{steady-analyt-inf-lambdaG-def}
\lambda_{G} = \sqrt{\frac{\tau \eta v_F^2}{w_0}}
\end{equation}
is the Gurzhi length, which characterizes the fluid velocity profile given in Eq.~(\ref{steady-analyt-ux-sol}).
The terms in the curly brackets of Eq.~(\ref{steady-analyt-ux-sol}) describe a driving force and the fluid velocity profile is defined by the terms in the square brackets. As one can see, the Poiseuille (parabolic-like) profile of the electric current is reproduced at $\lambda_{G}/L_y \gtrsim1$. In the opposite limit $\lambda_{G}/L_y \ll 1$, the Ohmic profile with almost uniform velocity is realized.

\begin{figure*}[ht]
\centering
\subfigure[]{\includegraphics[width=0.31\textwidth]{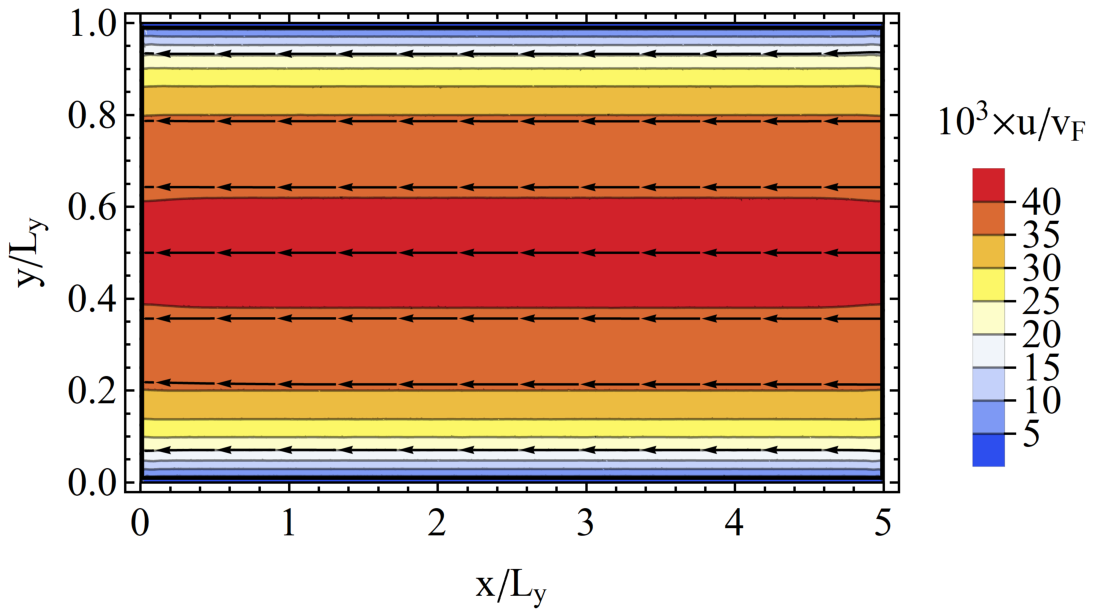}}
\hspace{0.02\textwidth}
\subfigure[]{\includegraphics[width=0.31\textwidth]{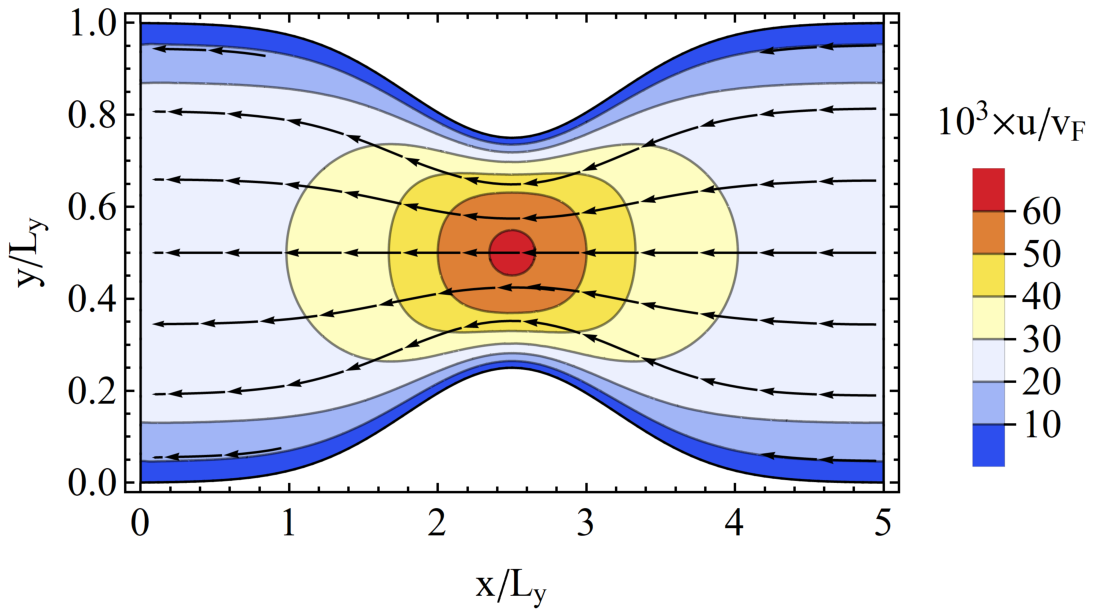}}
\hspace{0.02\textwidth}
\subfigure[]{\includegraphics[width=0.31\textwidth]{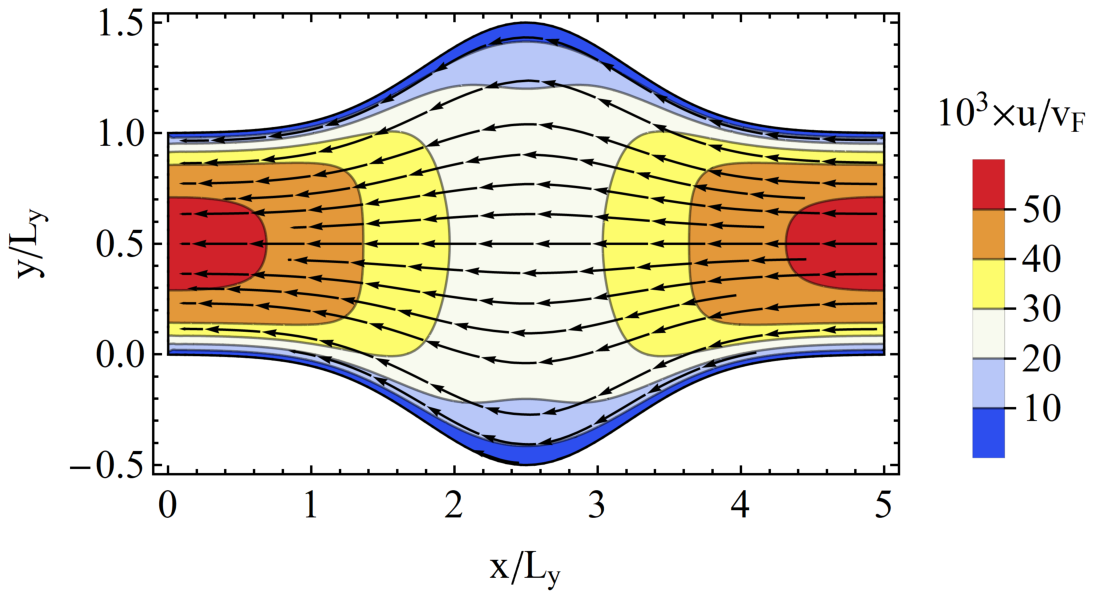}}
\caption{The profile of fluid velocity $\mathbf{u}(x,y)$ in the straight channel (a), nozzle (b), and cavity (c) geometries. Numerical parameters defined in Sec.~\ref{sec:model-eqs-BC} are
used.
}
\label{fig:steady-num-u-2D}
\end{figure*}

Next, we provide the numerical results for a 3D channel and discuss their key features for the model defined in Sec.~\ref{sec:model}. We focus on the hallmark hydrodynamic characteristics of fluids, i.e., the electron fluid velocity $\mathbf{u}(x,y)$. Notice that in the configuration at hand, the fluid velocity does not depend on the $z$-coordinate. Therefore, qualitatively similar profiles are realized for 2D ribbons. The nontrivial profile of the fluid velocity is directly reflected in the electric current density. We present the results for $\mathbf{u}(x,y)$ in Fig.~\ref{fig:steady-num-u-2D}. The velocity in the middle of the channel ($x=L_x/2$) shows the expected Poiseuille profile, which agrees with the analytical results in Eq.~(\ref{steady-analyt-ux-sol}). The fluid velocity becomes nonuniform along the $x$ direction in nozzles and cavities: it  increases in nozzles and decreases in cavities. Therefore, transport regimes with different flow velocities could be accessed by carefully engineering the sample.

\section{Time-dependent flow}
\label{sec:dynamic}

Let us proceed to the case of a time-dependent flow driven by an oscillating voltage; see Eq.~(\ref{model-phi-BC}) for the corresponding boundary conditions.

\subsection{Analytical model}
\label{sec:dynamic-analyt}

We start with an analytically solvable model of a straight channel infinite along the $x$ direction. For the sake of simplicity, we assume that the charge and energy densities remain constant, $n=n_0$ and $\epsilon=\epsilon_0$. (The former condition is not needed in 2D where the gradual channel approximation is used.) These approximations are due to strong screening and weak heating effects, respectively. Then the system of Eqs.~(\ref{model-Eq1})--(\ref{model-Eq4}) reduces to
\begin{eqnarray}
\label{dynamic-analyt-inf-Eq1-y}
&&\partial_{t}u_y(t,x,y)  -\frac{\eta v_F^2}{w_0} \Delta u_y(t,x,y) +\frac{u_y(t,x,y)}{\tau}= 0,\\
\label{dynamic-analyt-inf-Eq1-x}
&&\partial_{t}u_x(t,x,y)  -\frac{\eta v_F^2}{w_0} \Delta u_x(t,x,y) +\frac{u_x(t,x,y)}{\tau}=  -\frac{ev_F^2n_0E_{0}}{w_0} f(t,y).
\end{eqnarray}
We use the no-slip boundary conditions for $u_{x}(t,x,y)$ and $u_{y}(t,x,y)$ together with the initial conditions $u_{y}(t=0,x,y)=u_{x}(t=0,x,y)=0$. In the 2D case and for the gradual channel approximation, the pressure gradient should be taken into account. Then, the driving force $E_{0} f(t,y)$ on the right-hand side of Eq.~(\ref{dynamic-analyt-inf-Eq1-x}) reads as $E_{0} f(t,y) \left\{1-2C/\left[e^2 (\partial_{\mu} n)\right]\right\}$. This modification does not lead to any qualitative or, for the parameters at hand, even noticeable quantitative changes.

The solution to Eq.~(\ref{dynamic-analyt-inf-Eq1-y}) is trivial, i.e., $u_{y}(t,x,y)=0$. Because there is no dependence on the $x$-coordinate in the model at hand, the solution for $u_{x}(t,y)$ reads as
\begin{equation}
\label{dynamic-analyt-inf-ux-sol}
u_{x}(t,y) = -\frac{ev_F^2n_0E_{0}}{w_0} \int_{0}^{\infty}dt^{\prime} \int_{0}^{L_y}dy^{\prime} G\left(t-t^{\prime}; y,y^{\prime}\right) f(t^{\prime},y^{\prime}),
\end{equation}
where
\begin{equation}
\label{dynamic-analyt-inf-G-def}
G\left(t-t^{\prime}; y,y^{\prime}\right) = \frac{2}{L_y} \sum_{k=1}^{\infty}  e^{-C_k \left(t-t^{\prime}\right)/\tau} \theta\left(t-t^{\prime}\right) \sin{\left(\frac{\pi k y^{\prime}}{L_y}\right)} \sin{\left(\frac{\pi k y}{L_y}\right)}
\end{equation}
is the Green function and $C_k=1+\left(\lambda_G\pi k/L_y\right)^2$. Calculating the integrals over $t^{\prime}$ and $y^{\prime}$ in Eq.~(\ref{dynamic-analyt-inf-ux-sol}), the solution for the fluid velocity reads as
\begin{eqnarray}
\label{dynamic-analyt-inf-ux-sol-fin}
u_{x}(t,y) &=& -\frac{ev_F^2n_0E_{0} \tau}{w_0} \sum_{k=1,3,5,...} \frac{4}{\pi k} \sin{\left(\frac{\pi k y}{L_y}\right)} \frac{2\pi\nu \tau \sin{(2\pi\nu t)} +C_k \left[\cos{(2\pi\nu t)} -e^{-C_k t/\tau}\right]}{(2\pi\nu \tau)^2 + C_k^2} \nonumber\\
&\stackrel{t\gg\tau}{\approx}&
-\frac{ev_F^2n_0E_{0} \tau}{w_0} \sum_{k=1,3,5,...} \frac{4}{\pi k} \sin{\left(\frac{\pi k y}{L_y}\right)} \frac{2\pi\nu \tau \sin{(2\pi\nu t)} +C_k \cos{(2\pi\nu t)}}{(2\pi\nu \tau)^2 + C_k^2}.
\end{eqnarray}
As one can already see from the above equation, the spatial dependence of the fluid velocity $u_{x}(t,y)$ could be rather nontrivial for a nonzero frequency of the driving force. This is related to the interplay of the viscous and dissipative effects on the one hand and the acceleration due to the time-dependent external force on the other hand.

\begin{figure*}[ht]
\centering
\subfigure[]{\includegraphics[width=0.35\textwidth]{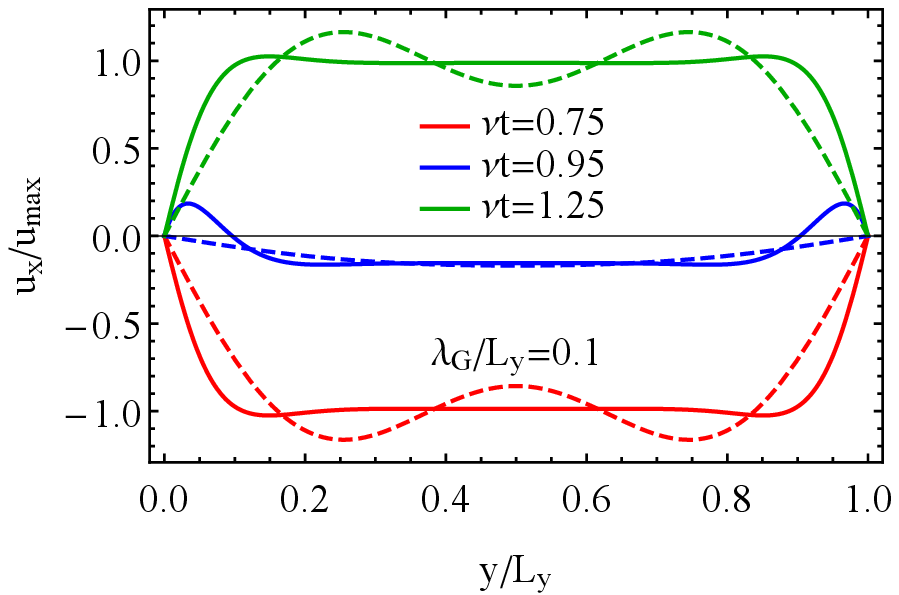}}
\hspace{0.05\textwidth}
\subfigure[]{\includegraphics[width=0.35\textwidth]{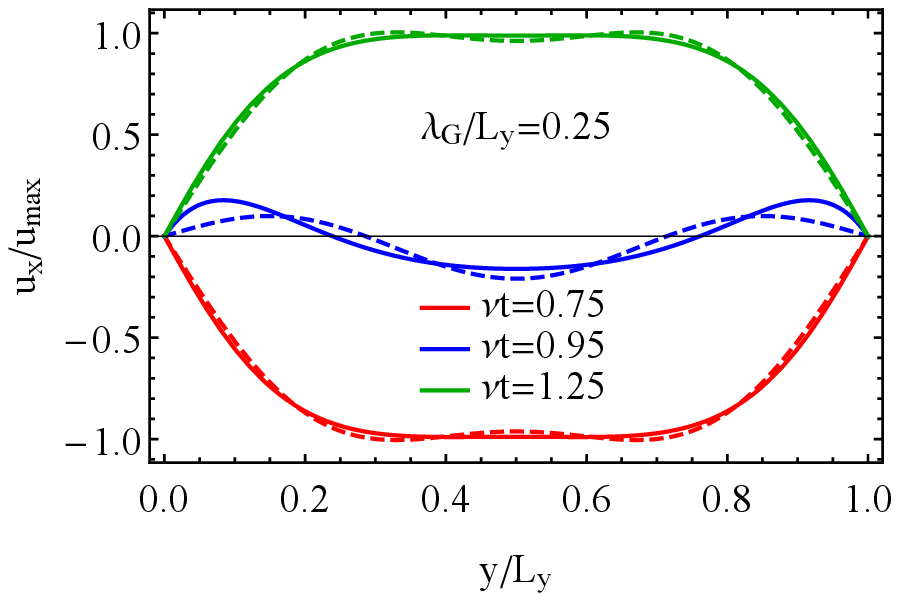}}
\caption{The fluid velocity profiles for a few values of $\nu t$ at $\lambda_{G}=0.1\, L_y$ (a) and $\lambda_{G}=0.25\, L_y$ (b). Solid and dashed lines correspond to $500$ and $2$ summands in Eq.~(\ref{dynamic-analyt-inf-ux-sol-fin}), respectively. Here, $u_{\rm max}$ is the maximum value of the fluid velocity across the channel during the period of the oscillating driving force.
}
\label{fig:dynamic-analyt-u}
\end{figure*}

We present the fluid velocity profiles in the transition regime in Fig.~\ref{fig:dynamic-analyt-u} for a few values of $t$. First, we notice that there is a phase shift between the driving force and the fluid velocity due to the acceleration term $\partial_t u_x$ [see also the term $2\pi \nu \tau\sin{(2\pi \nu t)}$ in Eq.~(\ref{dynamic-analyt-inf-ux-sol-fin})]. Indeed, while the driving force, according to Eq.~(\ref{model-f-BC}), vanishes at $\nu t\approx(2l+1)/4$ with $l=0,1,2,3,\ldots$, the velocity in a clean system (large $\nu \tau$) passes through zero for $\nu t\approx l/2$ in the middle of the channel; see also Appendix~\ref{sec:app-u-additional} where the time dependence of the fluid velocity is considered.

Another key feature of the result in Eq.~(\ref{dynamic-analyt-inf-ux-sol-fin}) is the appearance of a \emph{double-peak} structure in the transition regime. First, backflows occur near the boundaries. After delay determined by $\nu \tau$ and $\lambda_{G}/L_y$, the velocity in the bulk also changes sign. The obtained results agree with those in Ref.~\cite{Moessner-Witkowski:2018} where a different representation for the solution was used, see also Appendix~\ref{sec:app-2}.

The reasons for the formation of the double-peak structure could be clearly seen from Eq.~(\ref{dynamic-analyt-inf-ux-sol-fin}). Indeed, the first term in the sum, i.e., at $k=1$, is responsible for the Poiseuille fluid profile where $u_x(t,y)$ vanishes only at the boundaries. The second nonvanishing term has two zeros inside the sample as a function of $y$ that explains the formation of two peaks if other summands can be neglected. This is indeed the case in the viscous regime where $\lambda_{G}/L_y\gtrsim1/\pi$. In this regime, the terms with $C_k$ allow for a quick convergence of the sum in Eq.~(\ref{dynamic-analyt-inf-ux-sol-fin}) where only the first two terms are usually relevant. In the opposite limit, $\lambda_{G}/L_y\lesssim1/\pi$, $C_k$ is large and the sum converges slowly, leading to a flat Ohmic-like profile. Finally, comparing the solid and dashed lines in Fig.~\ref{fig:dynamic-analyt-u}(b), one can see that the velocity profile is well reproduced by taking into account only the first two nonvanishing terms in the sum in Eq.~(\ref{dynamic-analyt-inf-ux-sol-fin}).

For a nonzero $\nu \tau$, the convergence of the sum depends on $t$. Indeed, while at $\nu t\approx l/2$ the summand is $\propto 1/k^3$, for $\nu t\approx (2l+1)/4$, it decays as $\propto 1/k^5$. This explains why the fluid velocity profile changes its form with time.

Let us clarify the physics behind the nontrivial dynamics of the fluid velocity. There are two dimensionless parameters that determine the profile of the flow: (i) $\nu \tau$ and (ii) $\lambda_{G}/L_y$. The first parameter quantifies the inertial properties of the flow and the second parameter is related to fluid viscosity. We present the phase diagram of the system in the plane of these parameters in Fig.~\ref{fig:dynamic-analyt-phase} by using the curvature of the fluid flow (see Appendix~\ref{sec:app-u-additional} for its definition) and the phase shift between the fluid velocity and the driving force as guides. We identify three main regimes. The first regime is the Ohmic one where the current profile is flat in the middle of the channel. It occurs for $\nu \tau\ll1$ and $\lambda_{G}/L_y\ll1$. If the frequency is large enough and the viscosity is small but nonnegligible, the double peaks might occur near boundaries. However, they are hardly discernible on top of abrupt changes in the fluid profile. The second regime is the viscous regime of the fluid flow. It requires large viscosity $\lambda_{G}/L_y\gtrsim1$ and can be easily identified via the curvature of the velocity profile, see Fig.~\ref{fig:dynamic-analyt-phase}(a). Finally, the third regime has a noticeable phase shift between the driving force and the fluid velocity, see Fig.~\ref{fig:dynamic-analyt-phase}(b). For large $\nu\tau$, the driving force and the fluid velocity can oscillate in the antiphase, i.e., the phase shift in the middle of the channel reaches $\pi/2$. This regime is characterized by a double-peak profile of the fluid velocity. While the main requirement for the realization of the peaks is $\nu\tau\gtrsim1$, they are well manifested only if the fluid velocity has a well-pronounced Poiseuille profile, i.e., $\lambda_{G}/L_y$ should be nonnegligible.

\begin{figure*}[ht]
\centering
\subfigure[]{\includegraphics[width=0.35\textwidth]{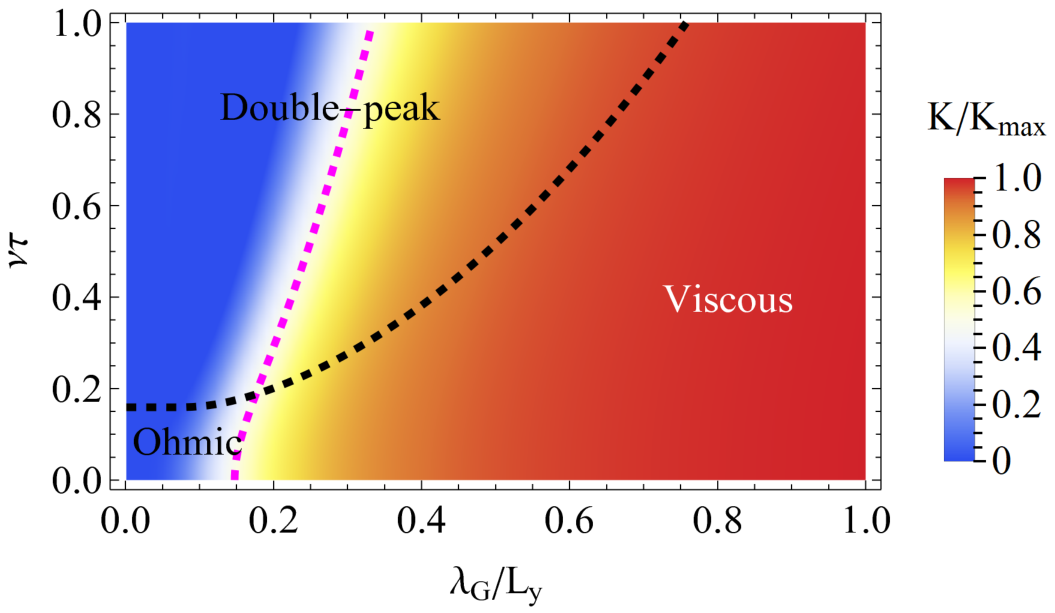}}
\hspace{0.05\textwidth}
\subfigure[]{\includegraphics[width=0.35\textwidth]{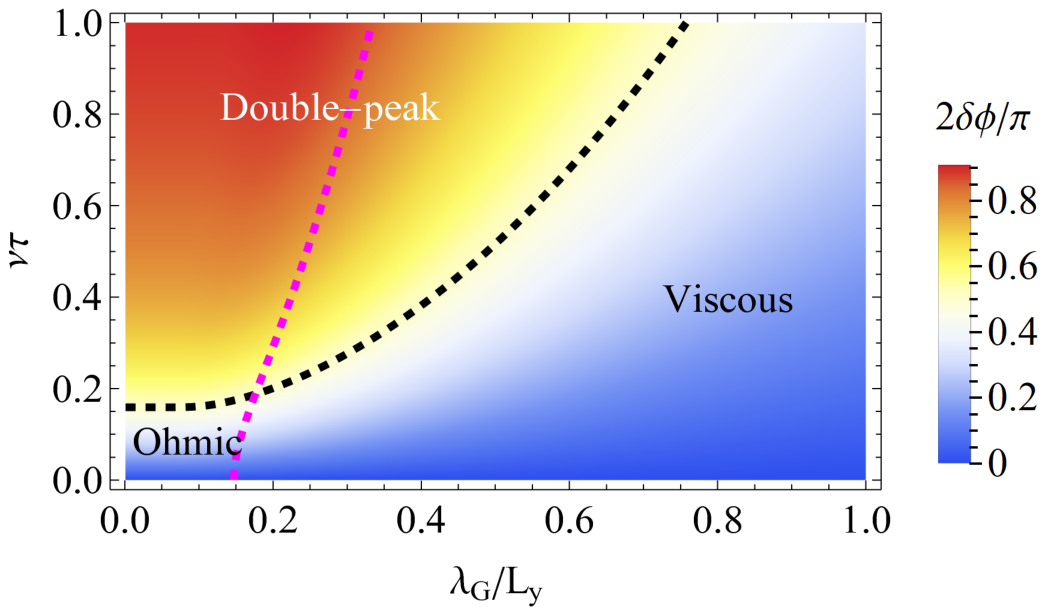}}
\caption{The phase diagram of the system in the plane of $\nu \tau$ and $\lambda_{G}/L_y$. We show the normalized curvature of the fluid flow in the middle of the channel in (a) and the phase difference $\delta \phi$ between the fluid velocity and the driving force in (b). The curvature of the fluid profile $K$ is defined in Appendix~\ref{sec:app-u-additional}. The phase shift is measured in the middle of the channel $y=L_y/2$. Black and magenta dashed lines correspond to $\delta \phi=\pi/4$ and $K=K_{\rm max}/2$, respectively.
}
\label{fig:dynamic-analyt-phase}
\end{figure*}

Finally, let us comment on the role of the boundary conditions and the time dependence of the driving force. As we show in Appendix~\ref{sec:app-2}, the boundary conditions for the fluid velocity play an important role in the formation of double peaks. In particular, there are no peaks for the free-surface boundary conditions where the fluid velocity profile remains flat. Still, due to the dominant role of the inertial term $\sim \nu \tau$, the oscillations of the velocity and the driving force are shifted in phase. As for the time profile of the driving force, the formation of the double-peak structure does not rely on the specific harmonic form used in Eq.~(\ref{dynamic-analyt-inf-ux-sol-fin}) and should occur as long as the driving force changes its sign. This is confirmed by the results for a wavelet profile of the driving force given in Appendix~\ref{sec:app-2}.

\subsection{Numerical results}
\label{sec:dynamic-num}

In this section, to support the analytical findings in Sec.~\ref{sec:dynamic-analyt} and investigate the stability of the double-peak profile with respect to the shape of the channel, we present numerical results for dynamical electron fluid flow in the straight channel, nozzle, and cavity geometries. (We consider a 3D channel for example.) We demonstrate that while the details of the velocity distribution noticeably depend on the geometry of the channel, the double-peak profile of the fluid velocity remains in the transient regime.

\begin{figure*}[ht]
\centering
\subfigure[]{\includegraphics[width=0.31\textwidth]{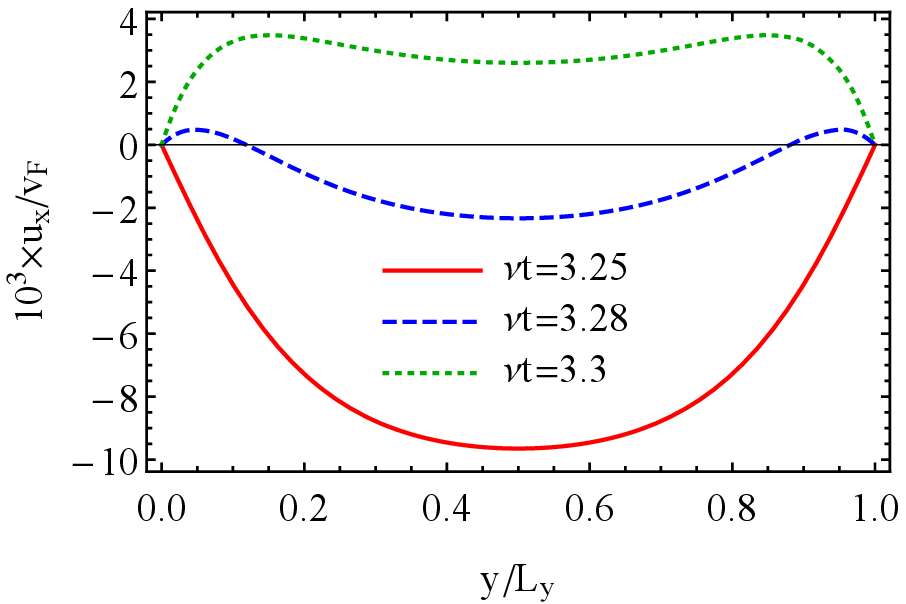}}
\hspace{0.02\textwidth}
\subfigure[]{\includegraphics[width=0.31\textwidth]{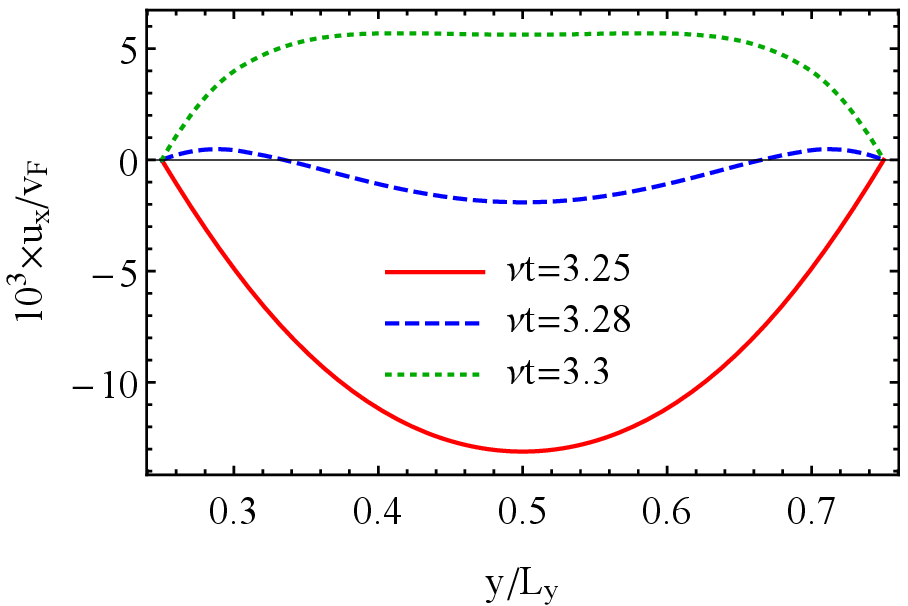}}
\hspace{0.02\textwidth}
\subfigure[]{\includegraphics[width=0.31\textwidth]{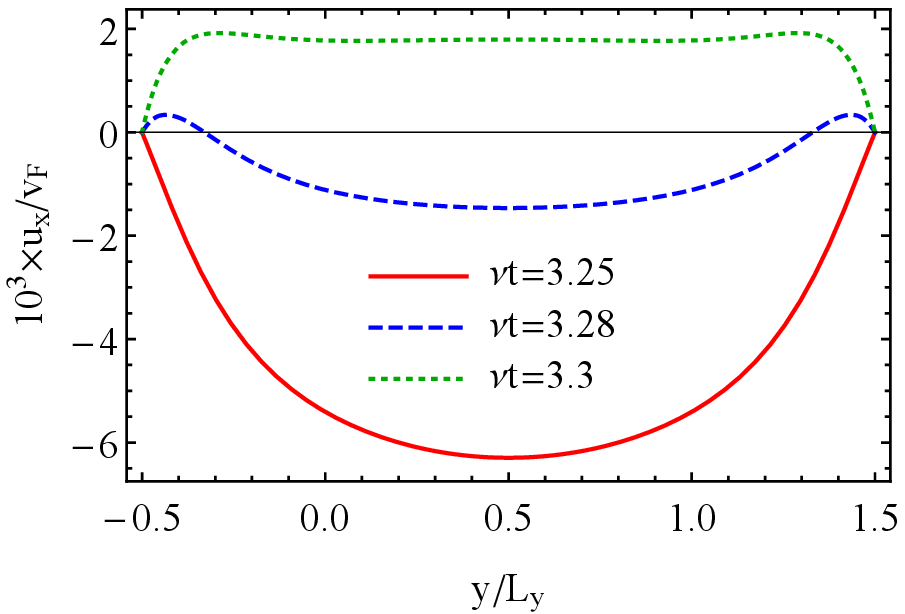}}
\caption{The fluid velocity $u_x(t,x,y)$ at $x=L_x/2$ for $\nu t=3.25$ (red solid line), $\nu t=3.28$ (blue dashed line), and $\nu t=3.3$ (green dotted line). Other parameters are defined in Sec.~\ref{sec:model-eqs-BC}. We use the channel (a), nozzle (b), and cavity geometries (c).
}
\label{fig:dynamic-num-all-uz-1D}
\end{figure*}

We focus on the most interesting case of the transition regime where the driving force changes its sign. It occurs at $\nu t\approx(2l+1)/4$ with $l=0,1,2,3,\ldots$. As one can see from Fig.~\ref{fig:dynamic-num-all-uz-1D}, after a short delay determined by $\nu \tau$, the fluid velocity also starts to change its sign in the middle of the channel. The transition, is, however, rather nontrivial. First, backflows occur near the boundaries leading to a \emph{double-peak} structure (see the blue dashed lines in Fig.~\ref{fig:dynamic-num-all-uz-1D}). Then, the velocity in the bulk also changes sign (see the green dotted lines in Fig.~\ref{fig:dynamic-num-all-uz-1D}). Finally, the velocity profile deforms and the peaks flatten  ultimately leading to a reversed profile compared to that shown by the red lines in Fig.~\ref{fig:dynamic-num-all-uz-1D}. These results perfectly agree with the analytical considerations in Sec.~\ref{sec:dynamic-analyt}.

\begin{figure*}[ht]
\centering
\subfigure[]{\includegraphics[width=0.31\textwidth]{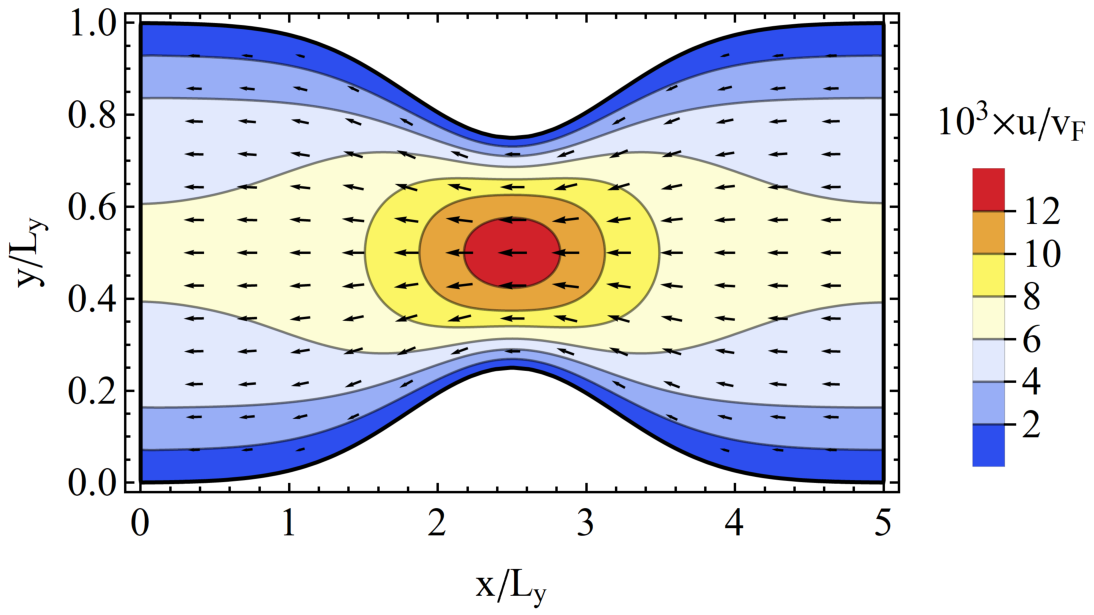}}
\hspace{0.02\textwidth}
\subfigure[]{\includegraphics[width=0.31\textwidth]{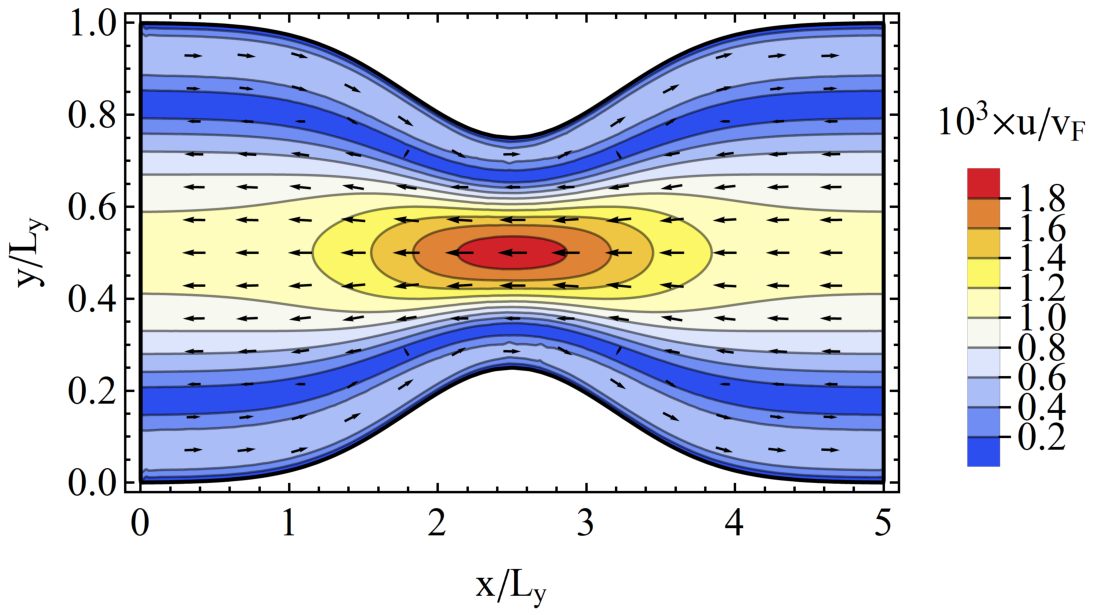}}
\hspace{0.02\textwidth}
\subfigure[]{\includegraphics[width=0.31\textwidth]{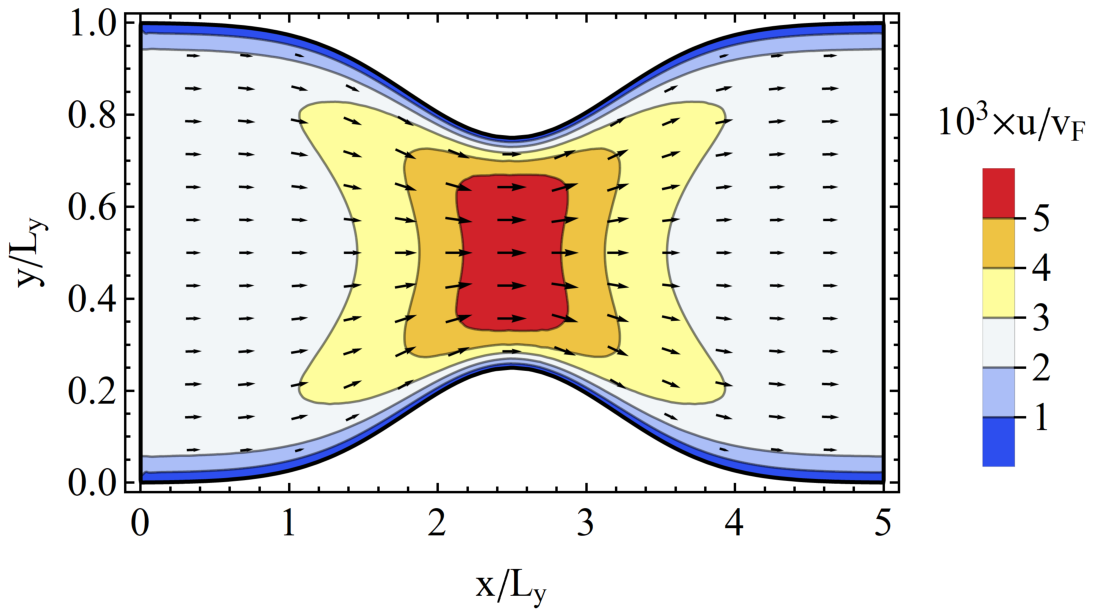}}
\caption{
The distribution of the fluid velocity $\mathbf{u}(t,x,y)$ in the nozzle geometry at $\nu t=3.25$ (a), $\nu t=3.28$ (b), and $\nu t=3.3$ (c). Other parameters are defined in Sec.~\ref{sec:model-eqs-BC}.
}
\label{fig:dynamic-num-nozzle-u-2D}
\end{figure*}

\begin{figure*}[ht]
\centering
\subfigure[]{\includegraphics[width=0.31\textwidth]{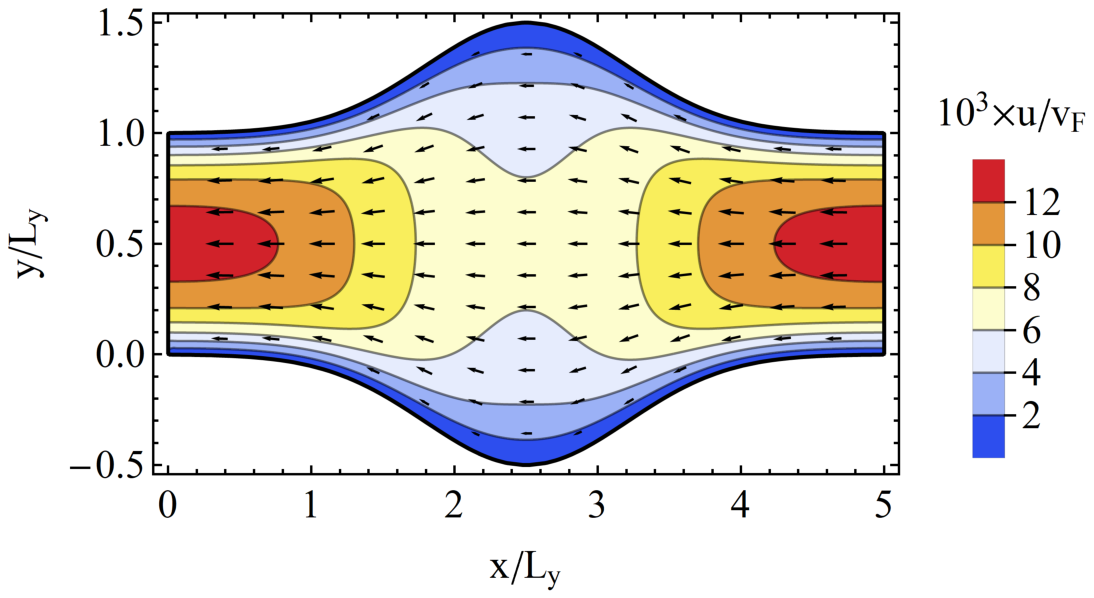}}
\hspace{0.02\textwidth}
\subfigure[]{\includegraphics[width=0.31\textwidth]{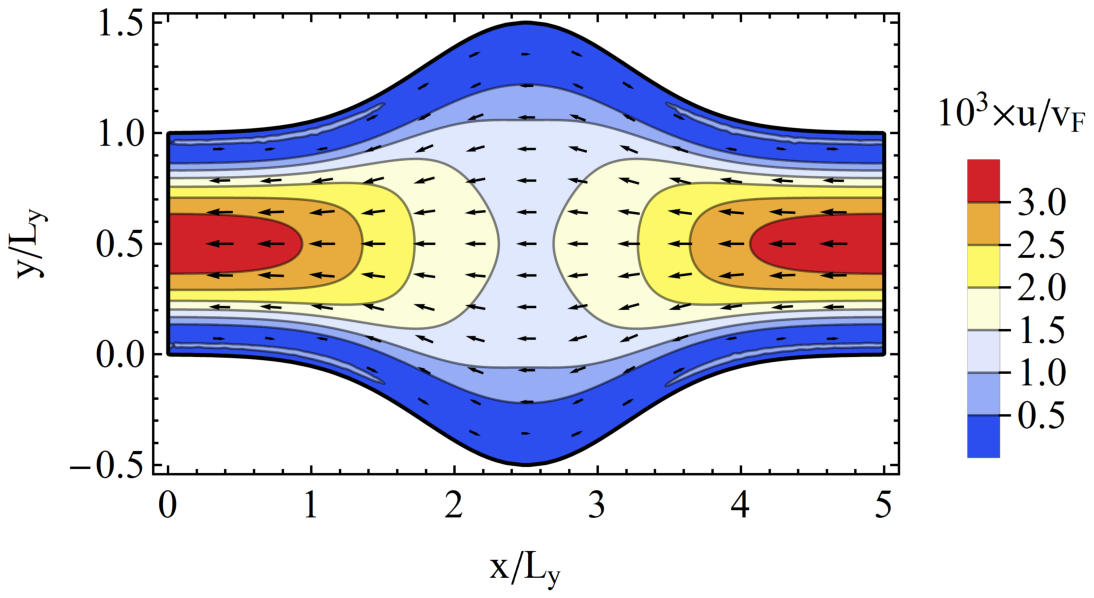}}
\hspace{0.02\textwidth}
\subfigure[]{\includegraphics[width=0.31\textwidth]{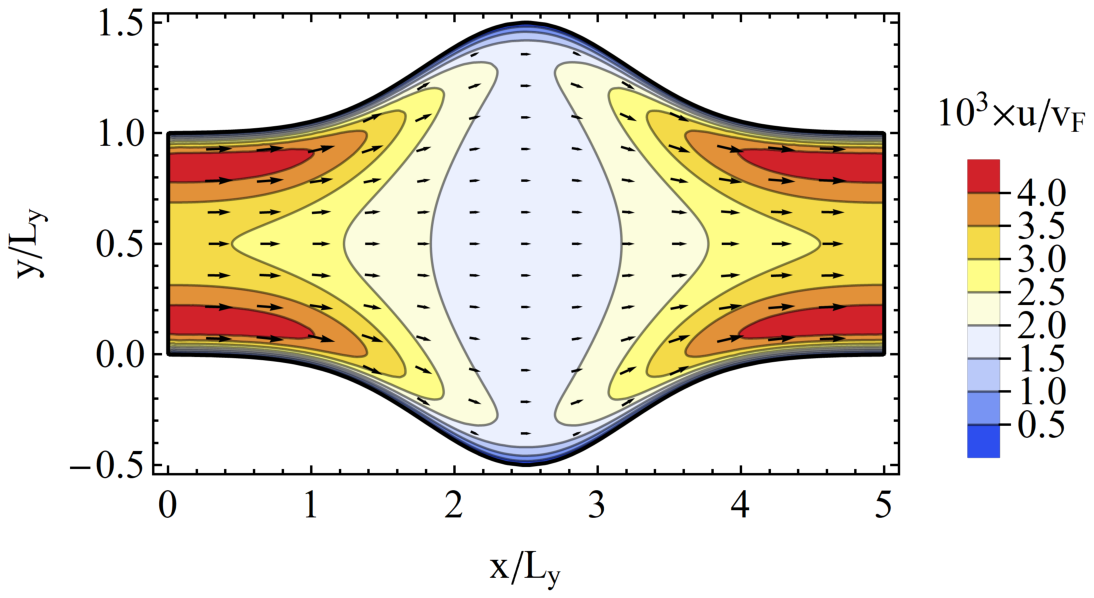}}
\caption{The distribution of the fluid velocity $\mathbf{u}(t,x,y)$ in the cavity geometry at $\nu t=3.25$ (a), $\nu t=3.28$ (b), and $\nu t=3.3$ (c). Other parameters are defined in Sec.~\ref{sec:model-eqs-BC}.
}
\label{fig:dynamic-num-cavity-u-2D}
\end{figure*}

The fluid flow in nozzle and cavity is visualized in Figs.~\ref{fig:dynamic-num-nozzle-u-2D} and \ref{fig:dynamic-num-cavity-u-2D}, respectively. A double-peak profile of the velocity can be seen in Figs.~\ref{fig:dynamic-num-nozzle-u-2D}(b) and \ref{fig:dynamic-num-cavity-u-2D}(b); see also Figs.~\ref{fig:dynamic-num-all-uz-1D}(b) and \ref{fig:dynamic-num-all-uz-1D}(c) for the velocity profile at $x=L_x/2$. Therefore, the formation of the double peaks in the transient regime is robust with respect to the channel geometry.

The overall distribution of the velocity is, as expected, modified. As in the case of the steady flow, the velocity increases in nozzles and diminishes in cavities. This can be used to effectively tune the flow velocity and access different transport regimes in the same sample. For example, by driving a large current through the sample it might even be possible to observe a transition from sub- to supersonic flow in nozzles~\cite{Moors-Kashuba:2019}.

\section{Stray magnetic field}
\label{sec:magnetic-field}

In this section, motivated by the recent studies of viscous flow via the QSM in Refs.~\cite{Ku-Walsworth:2019,Vool-Yacoby:2020-WTe2}, we show that the transient dynamics of pulsating viscous electron flow is manifested in stray magnetic fields generated by the electric current in the channel. To find the corresponding magnetic field, we employ the standard Maxwell equations
\begin{equation}
\label{MF-Maxwell-ee}
\bm{\nabla}\times\mathbf{B}(t,\mathbf{r}) = \frac{4\pi}{c}\mathbf{J}(t,\mathbf{r}) +\frac{1}{c}\partial_t\mathbf{E}(t,\mathbf{r}) \quad \mbox{and} \quad \bm{\nabla}\cdot\mathbf{B}(t,\mathbf{r}) = 0.
\end{equation}
By using the quasistatic approximation, introducing the vector potential $\mathbf{A}(t,\mathbf{r})$, and using the Coulomb gauge $\bm{\nabla}\cdot\mathbf{A}(t,\mathbf{r})=0$, we obtain
\begin{equation}
\label{MF-A-eq}
\Delta \mathbf{A}(t,\mathbf{r}) = \frac{4\pi}{c} \mathbf{J}(t,\mathbf{r}).
\end{equation}

The quasistatic approximation used in Eq.~(\ref{MF-A-eq}) is valid for sufficiently small frequencies such that $4\pi\left|\mathbf{J}(t,\mathbf{r})\right| \gg \left|\partial_t\mathbf{E}(t,\mathbf{r})\right|$. Let us estimate for which parameters this condition holds. At a fixed electric field, the largest value of the electric current can be estimated from Eq.~(\ref{steady-analyt-ux-sol}) at $y=L/2$ and $\lambda_{G}/L$. Then, we find that the displacement current is negligible in Eq.~(\ref{MF-Maxwell-ee}) at $w_0\omega/\left(4\pi e^2v_F^2n_0^2\tau\right) \ll1$ in a 3D setup. For the parameters presented at the end of Sec.~\ref{sec:model-eqs-BC}, this limits our consideration to a rather wide range of frequencies $\nu\ll 2.9\times10^{17}~\mbox{Hz}$.

Since the relation between the electric current and the stray magnetic field is unambiguous only in 2D or quasi-2D materials, we consider a long graphene ribbon where the width $L_y$ is much smaller than the length of the ribbon. The ribbon is located in the $x-y$ plane at $z=0$ and the gate is at $z=-L_g$. If the magnetic permeability of the gate is small, both vector potential and its derivatives are continuous at the surfaces of the gate~\cite{Landau:t8}. In addition, the vector potential should diminish for large $z$, i.e., $\mathbf{A}(t,\mathbf{r}_{\perp},z\to\pm\infty)\to\mathbf{0}$. Then, by using the fact that the current is localized in the ribbon, $\mathbf{J}(t,\mathbf{r})=\mathbf{J}(t,\mathbf{r}_{\perp})\delta(z)$, and performing the Fourier transform in the in-plane coordinates, we obtain
\begin{equation}
\label{MF-A-sol}
\mathbf{A}(t,\mathbf{q},z)= -\frac{2\pi}{cq} \mathbf{J}(t,\mathbf{q}) e^{-q|z|},
\end{equation}
where
\begin{equation}
\label{MF-Fourier-J}
\mathbf{J}(t,\mathbf{q}) = \int d^2\mathbf{r}_{\perp} e^{-i\mathbf{q}\cdot\mathbf{r}_{\perp}} \mathbf{J}(t,\mathbf{r}_{\perp}).
\end{equation}

We find that the components of the magnetic field in the coordinate space are
\begin{eqnarray}
\label{MF-B-x-sol-r-no-gate}
B_x(t,\mathbf{r}_{\perp},z) &=& -\frac{2\pi}{c} \int \frac{d^2\mathbf{q}}{(2\pi)^2} e^{i\mathbf{q}\cdot \mathbf{r}_{\perp}} J_y(t,\mathbf{q}) \sign{z} e^{-q|z|},\\
\label{MF-B-y-sol-r-no-gate}
B_y(t,\mathbf{r}_{\perp},z) &=&  \frac{2\pi}{c}  \int \frac{d^2\mathbf{q}}{(2\pi)^2} e^{i\mathbf{q}\cdot \mathbf{r}_{\perp}}J_x(t,\mathbf{q}) \sign{z} e^{-q|z|},\\
\label{MF-B-z-sol-r-no-gate}
B_z(t,\mathbf{r}_{\perp},z) &=& -i\frac{2\pi}{c}  \int \frac{d^2\mathbf{q}}{(2\pi)^2} e^{i\mathbf{q}\cdot \mathbf{r}_{\perp}} \frac{\left[\mathbf{q}\times \mathbf{J}(t,\mathbf{q})\right]_z}{q} e^{-q|z|}.
\end{eqnarray}
In the case of a long straight ribbon, $J_y(t,\mathbf{r}_{\perp})=0$, and there is no dependence on the $x$-coordinate. Therefore, the only nontrivial components of the magnetic field are $B_y(t,y,z)$ and $B_z(t,y,z)$ given in Eqs.~(\ref{MF-B-y-sol-r-no-gate}) and (\ref{MF-B-z-sol-r-no-gate}). In addition, since $\mathbf{J}(t,\mathbf{q})= 2\pi\delta(q_x) \mathbf{J}(t,q_y)$, the integral over $q_x$ can be trivially taken.

\begin{figure*}[ht]
\centering
\includegraphics[width=0.35\textwidth]{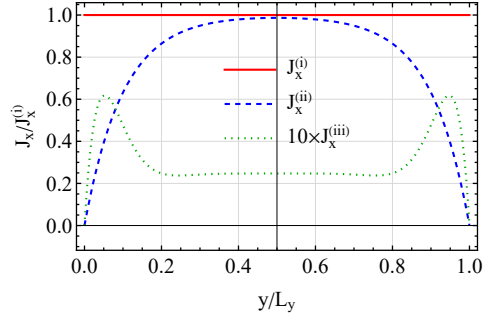}
\caption{The normalized electric current profiles that are used for calculating the magnetic field distribution. The red solid line corresponds to the Ohmic profile $J_{x}^{(i)}$, the blue dashed line describes the Poiseuille profile $J_{x}^{(ii)}$, and the green dotted line shows the double-peak profile $J_{x}^{(iii)}$. We use $\nu t=3$, $\nu \tau =1$, and $\lambda_G/L_g=0.1$ in Eq.~(\ref{dynamic-analyt-inf-ux-sol-fin}) for the double-peak profile and Eq.~(\ref{steady-analyt-ux-sol}) at $\partial_xT=0$ and $\lambda_G/L_g=0.1$ for the Poiseuille one.
}
\label{fig:MF-profiles}
\end{figure*}

In what follows, we present the result for the magnetic field for three different profiles of the electric current: (i) Ohmic with $J_x(t,y)= e^2 v_F^2 n_0^2 \tau E_0/w_0$, (ii) Poiseuille $J_x(t,y)=-en_0u_x(t,y)$ with $u_x(t,y)$ defined in Eq.~(\ref{steady-analyt-ux-sol}) at $\partial_xT=0$, and (iii) double-peak, where the velocity is given in Eq.~(\ref{dynamic-analyt-inf-ux-sol-fin}). We use $\nu t=3$, $\nu\tau=1$, and $\lambda_G/L_g=0.1$ for the double-peak profile. The electric current profiles are presented in Fig.~\ref{fig:MF-profiles}. Notice that since the double-peak profile is realized in the transient regime, its magnitude is typically smaller than that for a steady flow.

By using the typical numerical parameters for graphene [see Sec.~\ref{sec:model-eqs-BC}], we estimate the following reference value of the magnetic field:
\begin{equation}
\label{MF-num-estimate}
B_{*}=\frac{e^2v_F^2 n_0^2 \tau E_0}{cw_0} \approx 12\, \frac{E_0}{1~\mbox{V/cm}}~\mu\mbox{T},
\end{equation}
which is essentially the Ohmic current density divided by the speed of light. As we will show below, the typical values of generated magnetic fields are of the order of $B_{*}$. For example, the total current $I\approx1~\mu\mbox{A}$ considered in Ref.~\cite{Ku-Walsworth:2019} is equivalent to $E_0\sim1~\mbox{V/cm}$ and $B_{*} \sim 12~\mu\mbox{T}$ for the parameters used in our paper. The magnitude of such oscillating magnetic fields is within the reach of the QSM~\cite{Levine-Walsworth:2019-QSM,Barry-Walsworth:2019-QSM} and the scanning SQUID magnetometry~\cite{Cui-Moler:2017}. However, it could be experimentally challenging to probe these fields in view of their relatively high frequency. As for the restrictions imposed by the limited spatial resolution, they can be lifted by using larger samples.

\begin{figure*}[ht]
\centering
\subfigure[]{\includegraphics[width=0.35\textwidth]{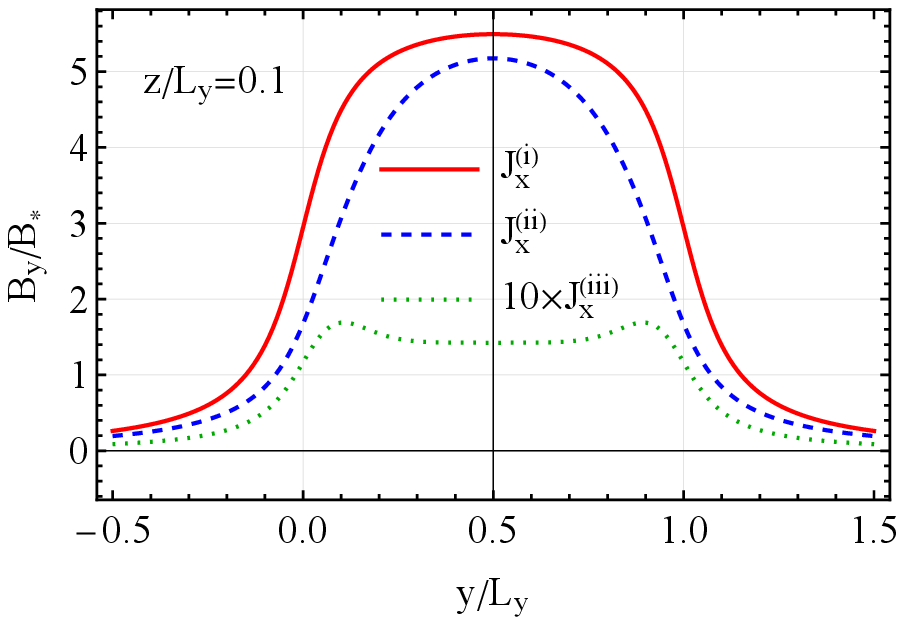}}
\hspace{0.05\textwidth}
\subfigure[]{\includegraphics[width=0.35\textwidth]{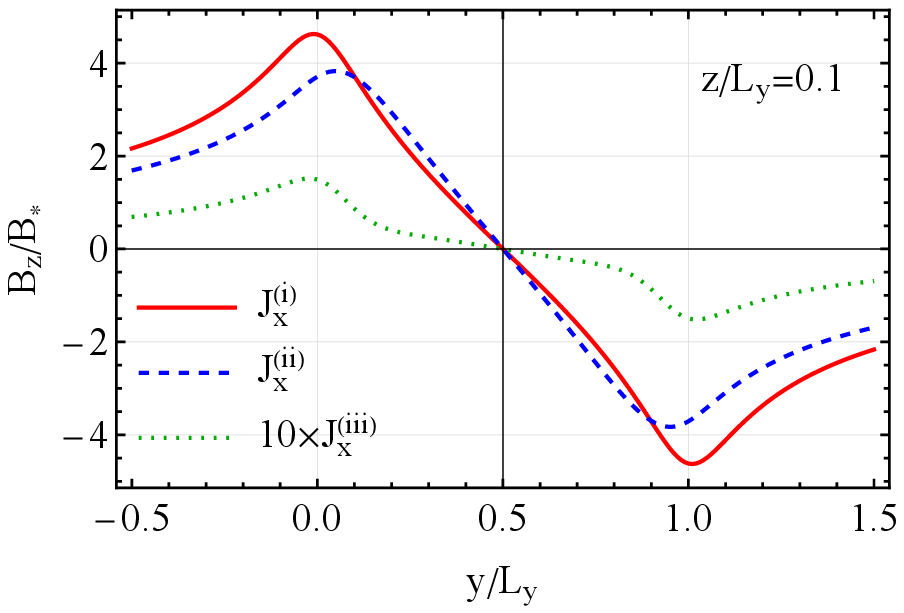}}
\caption{The normalized magnetic field profiles defined in Eqs.~(\ref{MF-B-y-sol-r-no-gate}) and (\ref{MF-B-z-sol-r-no-gate}) at $z/L_y=0.1$. Red solid, blue dashed, and green dotted lines correspond to the Ohmic $J_x^{(i)}$, Poiseuille $J_x^{(ii)}$, and double-peak $J_x^{(iii)}$ profiles of the electric current shown in Fig.~\ref{fig:MF-profiles}. We normalized the results by the reference value of the magnetic field defined in Eq.~(\ref{MF-num-estimate}). We use $\nu t=3$, $\nu \tau =1$, and $\lambda_G/L_g=0.1$ in Eq.~(\ref{dynamic-analyt-inf-ux-sol-fin}) for the double-peak profile of the electric current and Eq.~(\ref{steady-analyt-ux-sol}) at $\partial_xT=0$ and $\lambda_G/L_g=0.1$ for the Poiseuille one.
}
\label{fig:MF-By-Bz-all-no-gate}
\end{figure*}

\begin{figure*}[ht]
\centering
\subfigure[]{\includegraphics[width=0.35\textwidth]{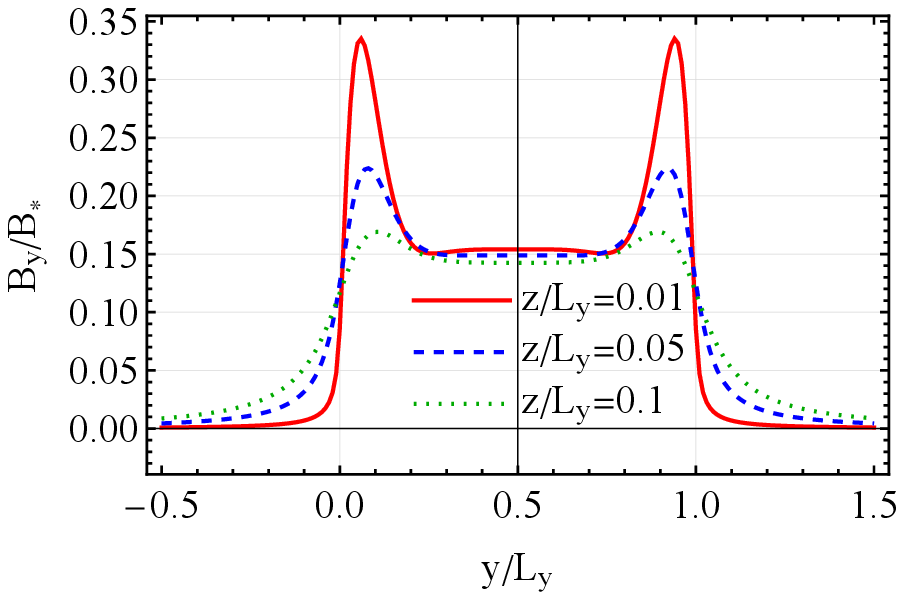}}
\hspace{0.05\textwidth}
\subfigure[]{\includegraphics[width=0.35\textwidth]{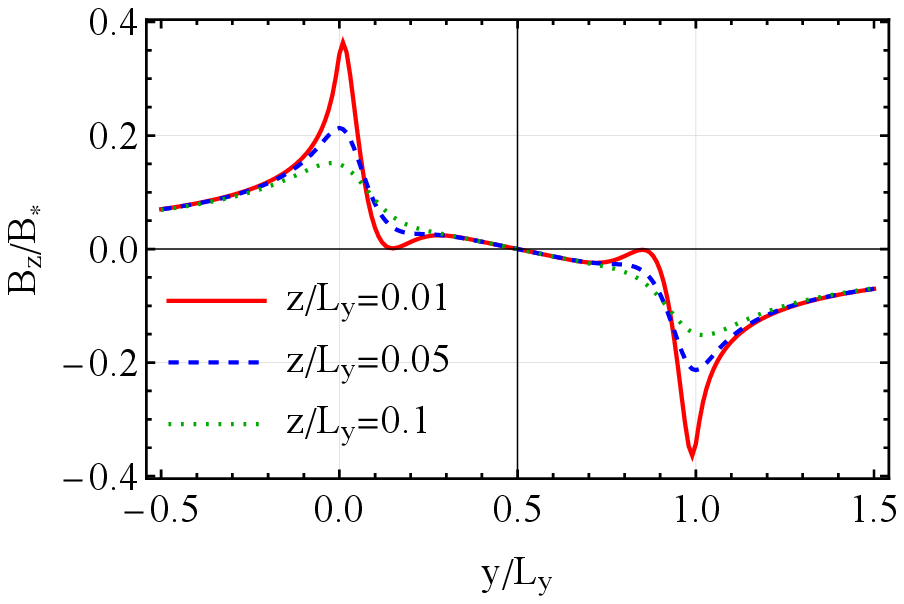}}
\caption{The normalized magnetic field profiles defined in Eqs.~(\ref{MF-B-y-sol-r-no-gate}) and (\ref{MF-B-z-sol-r-no-gate}) for the double-peak profile of the electric current for a few values of $z$: $z=0.01\, L_y$ (red solid line), $z=0.05\, L_y$ (blue dashed line), $z=0.1\, L_y$ (green dotted line). We normalized the results by the reference value of the magnetic field defined in Eq.~(\ref{MF-num-estimate}). We use $\nu t=3$, $\nu \tau =1$, and $\lambda_G/L_g=0.1$ in Eq.~(\ref{dynamic-analyt-inf-ux-sol-fin}).
}
\label{fig:MF-By-Bz-few-z-no-gate}
\end{figure*}

The obtained magnetic field profiles at fixed $z=0.1\, L_y$ are shown in Fig.~\ref{fig:MF-By-Bz-all-no-gate}. The magnetic field for a double-peak current profile at a few values of $z$ is presented in Fig.~\ref{fig:MF-By-Bz-few-z-no-gate}. The $y$ component of the field attains its maximal value in the middle of the ribbon and gradually diminishes away from it. The $z$ component changes its sign across the ribbon and its absolute value has maxima near the edges of the sample. As one can see, the field distribution for the Ohmic and Poiseuille profiles are qualitatively similar albeit have a slightly different curvature at $y=L_y/2$. Nevertheless, this subtle difference corresponds to qualitatively distinct electric current profiles.

The distribution of the magnetic field for the double-peak profile of the current differs from both Ohmic and Poiseuille currents. Among the most noticeable features, we notice a double-peak structure in $B_y$ with two symmetric peaks near the edges, see Figs.~\ref{fig:MF-By-Bz-all-no-gate}(a) and \ref{fig:MF-By-Bz-few-z-no-gate}(a). In addition, the peaks in the $z$ component of the field are steeper and might even be supplemented with additional extrema for $z\ll L_y$, see Fig.~\ref{fig:MF-By-Bz-few-z-no-gate}(b). These features could be used to pinpoint the double-peak profile of the electric current. It is worth noting, however, that since the double-peak structure appears in a transient regime, the magnitude of the electric current and, consequently, the generated magnetic field is smaller compared to that for the steady-state currents; see Fig.~\ref{fig:MF-By-Bz-all-no-gate} where we multiplied the results by $10$ for a time-dependent drive.

Before finalizing this section, let us briefly address the role of the gate in the magnetic field distribution. As we discussed before, the thin gate made of material with small magnetic permeability does not alter the vector potential and, consequently, the magnetic field distribution~\footnote{An oscillating magnetic field attenuates in a conductor with the penetration depth determined by the skin depth. Therefore, even at small magnetic permeability, the gate might distort the time-dependent magnetic field in its vicinity.}. If the gate has a large magnetic permeability, the boundary conditions for the vector potential play an important role. In particular, in the limit of infinitely large permeability, the spatial derivative of the vector potential vanishes at the gate, i.e., $\partial_z \mathbf{A}(t,\mathbf{r}_{\perp},z=-L_g)=\mathbf{0}$. The generalization of the above calculations is straightforward in this case, therefore, we do not present it here. We checked that while the magnetic field distribution near the gate (at $z<0$) noticeably changes, it remains qualitatively the same above the sample, i.e., at $z>0$, where it could be conveniently measured. Furthermore, the distortion of the magnetic field could be easily avoided by using gates with low magnetic permeability made from, e.g., graphite, copper, platinum, etc.

\section{Summary and discussions}
\label{sec:Summary}

We studied steady and transient flows of the electron liquid in channels (3D) and ribbons (2D) of different shapes, including the straight channel/ribbon, nozzle, and cavity geometries. Among the key findings of our paper is a nontrivial double-peak hydrodynamic profile for a transient flow that is manifested in a stray magnetic field. In addition, a phase diagram, criteria, and physical parameters needed for the realization of the double-peak structure are determined.
While we used a relativisticlike dispersion relation of quasiparticles, the obtained results are general and could be applied to materials with other energy dispersions.

In the case of a straight ribbon, it is demonstrated that, similarly to the current density, the in-plane magnetic field component perpendicular to the flow acquires a nontrivial profile with symmetric peaks near the boundaries, see Sec.~\ref{sec:magnetic-field}. The normal to the ribbon component of the field also shows nonmonotonic behavior with additional extrema compared to the magnetic field for both Ohmic (flat) and Poiseuille (parabolic) profiles of the current.

Let us discuss possible ways to measure the nontrivial profile of the stray magnetic field discussed above. There are only a few techniques to probe the profile of weak fields with high resolution. Among them, we mention the QSM and the scanning SQUID magnetometry. For example, using ensembles of nitrogen vacancy centers in diamond, the widefield microwave microscope with a few-micron spatial resolution was built and used to imagine a microwave magnetic field of various circuitry components~\cite{Horsley-Treutlein:2018}. As for the SQUID magnetometry, a scanning SQUID sampler with a $40$-ps time resolution and a micron-scale pickup loop for the detection of a magnetic field flux was designed and fabricated~\cite{Cui-Moler:2017}. Therefore, even with the available methods, the observation of the double-peak profile of the magnetic field predicted in our paper might be possible, at least in principle. The precision of these techniques might even further improve in the future. We notice, however, that the technical details of the experimental setups for observation of the magnetic field are outside the scope of our paper and will be reported elsewhere.

In addition to studying the stray magnetic field generated by a time-dependent electron flow, we also analyzed the conditions for the manifestation and stability of the double-peak profile. By representing the solution in a different form and identifying the dimensionless quantities that determine the dynamics of the electron fluid in the transient regime, we obtained the phase diagram of the system, see Sec.~\ref{sec:dynamic-analyt} and Fig.~\ref{fig:dynamic-analyt-phase}. In essence, the double peaks arise due to the interplay of the inertial and viscous properties. While the inertia leads to a phase difference between the driving force and the hydrodynamic current, the viscosity of the electron fluid allows for a double-peak profile to manifest.

To address the stability of the double-peak profile, we studied the role of the surface geometry, boundary conditions for the fluid velocity, and the time dependence of the driving force. While the velocity distribution is noticeably affected by the geometry of samples, the double-peak profile is robust as long as the fluid sticks to the boundaries (no-slip boundary conditions). As for the time dependence of the driving force, the key requirement is the presence of well-defined transition regions where the force changes its sign.

In addition, we demonstrated that nozzles and cavities can be used to locally enhance and reduce the velocity of the electron fluid, respectively. In particular, nozzles can be employed to reach high velocities and, potentially, achieve a nonlinear regime. It could be possible even to attain a choked flow regime~\cite{Dyakonov-Shur:1995}. Therefore, nozzles and cavities could be important in creating hydrodynamic electronics similar to conventional plumbing.

\begin{acknowledgments}
The Authors acknowledge useful communications with I.~A.~Shovkovy and P.~Sur\'{o}wka. P.O.S. acknowledges the support through the Yale Prize Postdoctoral Fellowship in Condensed Matter Theory.
\end{acknowledgments}

\appendix

\section{Thermodynamic relations}
\label{sec:app-1}

In this appendix, we present a few thermodynamic relations that are used in numerical estimates in the main text. In the case of a linear response, we use the following relations:
\begin{eqnarray}
\label{app-1-dn}
\bm{\nabla}n &=& (\partial_{\mu} n)\bm{\nabla} \mu +(\partial_{T} n) \bm{\nabla} T,\\
\label{app-1-deps}
\bm{\nabla}\epsilon &=& (\partial_{\mu} \epsilon) \bm{\nabla} \mu +(\partial_{T} \epsilon) \bm{\nabla} T.
\end{eqnarray}
Here, $n$ and $\epsilon$ are the electric charge and energy densities, $\mu$ is the chemical potential, and $T$ is temperature. We find it convenient for numerical calculations to rewrite the dynamical equations in Sec.~\ref{sec:model} in terms of $n$ and $\epsilon$. By using Eqs.~(\ref{app-1-dn}) and (\ref{app-1-deps}), we find
\begin{eqnarray}
\label{app-1-dmu}
\bm{\nabla}\mu &=& \frac{(\partial_{T}\epsilon)\bm{\nabla}n -(\partial_{T}n)\bm{\nabla}\epsilon}{(\partial_{\mu}n)(\partial_{T}\epsilon)-(\partial_{T}n)(\partial_{\mu}\epsilon)},\\
\label{app-1-dT}
\bm{\nabla}T &=& \frac{(\partial_{\mu}n)\bm{\nabla}\epsilon -(\partial_{\mu}\epsilon)\bm{\nabla}n}{(\partial_{\mu}n)(\partial_{T}\epsilon)-(\partial_{T}n)(\partial_{\mu}\epsilon)}.
\end{eqnarray}

Next, let us present the key thermodynamic variables. In a 3D relativisticlike case, the electron number density $n$ and the energy density $\epsilon$ read
\begin{equation}
\label{app-1-model-n}
n= N_{W}\frac{\mu \left(\mu^2 +\pi^2T^2\right)}{6\pi^2 v_F^3 \hbar^3}
\end{equation}
and
\begin{equation}
\label{app-1-model-epsilon}
\epsilon = N_{W}\frac{1}{8\pi^2 \hbar^3v_F^3} \left(\mu^4 +2\pi^2T^2\mu^2 +\frac{7\pi^4T^4}{15}\right),
\end{equation}
respectively. Here, $N_{W}$ is the number of Weyl nodes and $v_F$ is the Fermi velocity. The electron number density $n$ and the energy density $\epsilon$ for 2D relativisticlike spectrum are
\begin{equation}
\label{app-1-model-n-2D}
n= -N_g\frac{T^2}{2\pi v_F^2 \hbar^2} \left[\mbox{Li}_{2}\left(-e^{\mu/T}\right) -\mbox{Li}_{2}\left(-e^{-\mu/T}\right)\right]
\end{equation}
and
\begin{equation}
\label{app-1-model-eps-2D}
\epsilon = -N_g \frac{T^3}{\pi v_F^2 \hbar^2} \left[\mbox{Li}_{3}\left(-e^{\mu/T}\right) +\mbox{Li}_{3}\left(-e^{-\mu/T}\right)\right],
\end{equation}
respectively. Here, $N_g=4$ accounts for the valley and spin degeneracy in graphene.

\section{Fluid flow velocity}
\label{sec:app-u-additional}

\begin{figure*}[ht]
\centering
\subfigure[]{\includegraphics[width=0.31\textwidth]{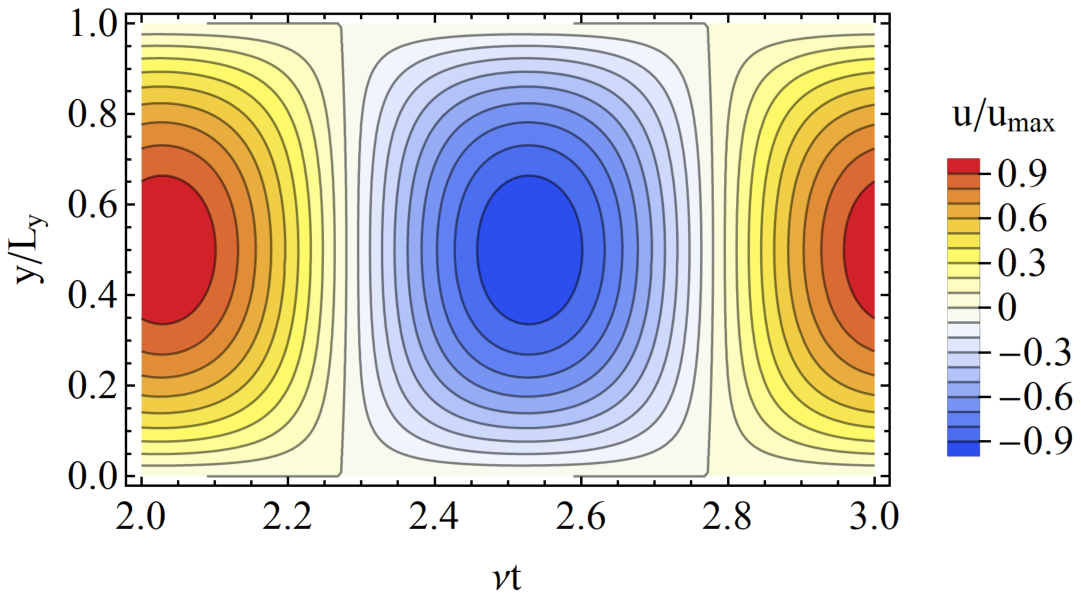}}
\hspace{0.02\textwidth}
\subfigure[]{\includegraphics[width=0.31\textwidth]{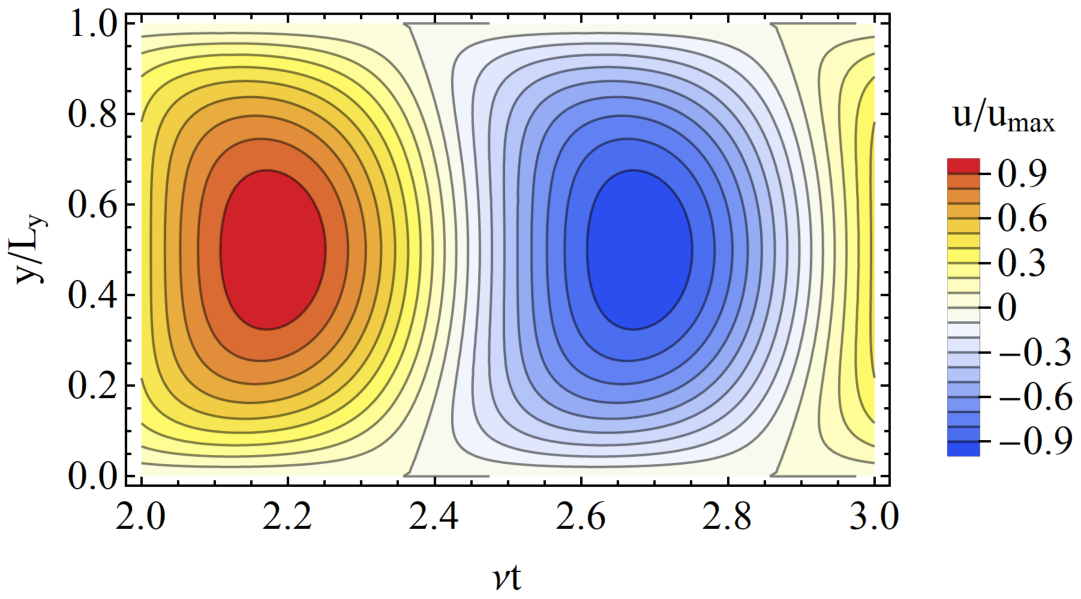}}
\hspace{0.02\textwidth}
\subfigure[]{\includegraphics[width=0.31\textwidth]{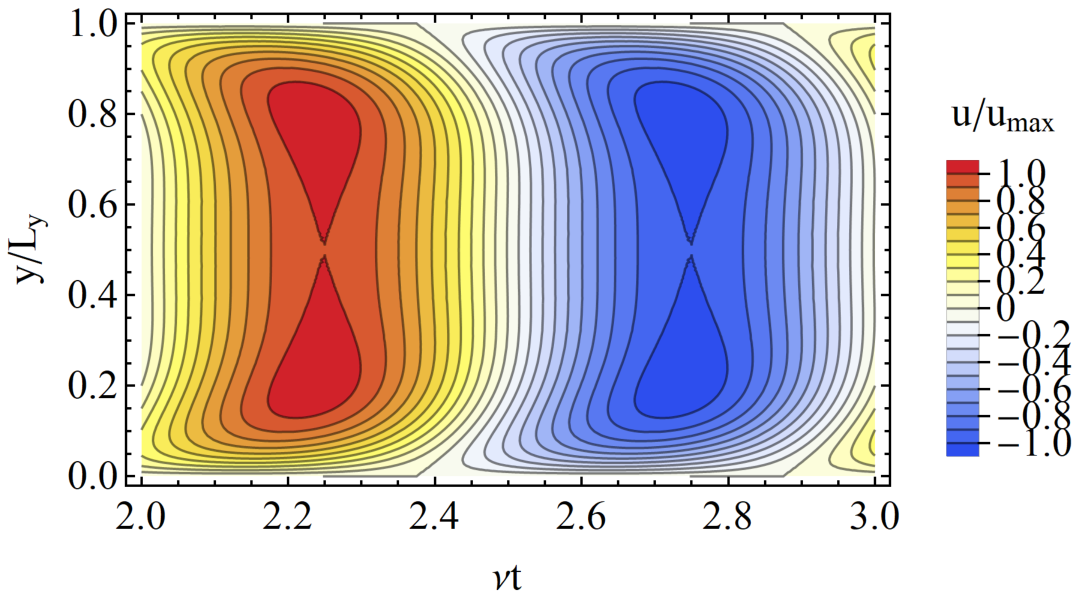}}
\caption{The normalized fluid velocity $\mathbf{u}(t,x,y)$ defined in Eq.~(\ref{dynamic-analyt-inf-ux-sol-fin}) for $\nu\tau=0.1$ (a), $\nu\tau=1$ (b), and $\nu\tau=10$ (c). In all panels, we fixed $\lambda_{G}/L_y=0.5$. In addition, $u_{\rm max}$ is the maximum value of the fluid velocity across the channel during the period of the oscillating driving force.
}
\label{fig:dynamic-analyt-ux-nu}
\end{figure*}

\begin{figure*}[ht]
\centering
\subfigure[]{\includegraphics[width=0.31\textwidth]{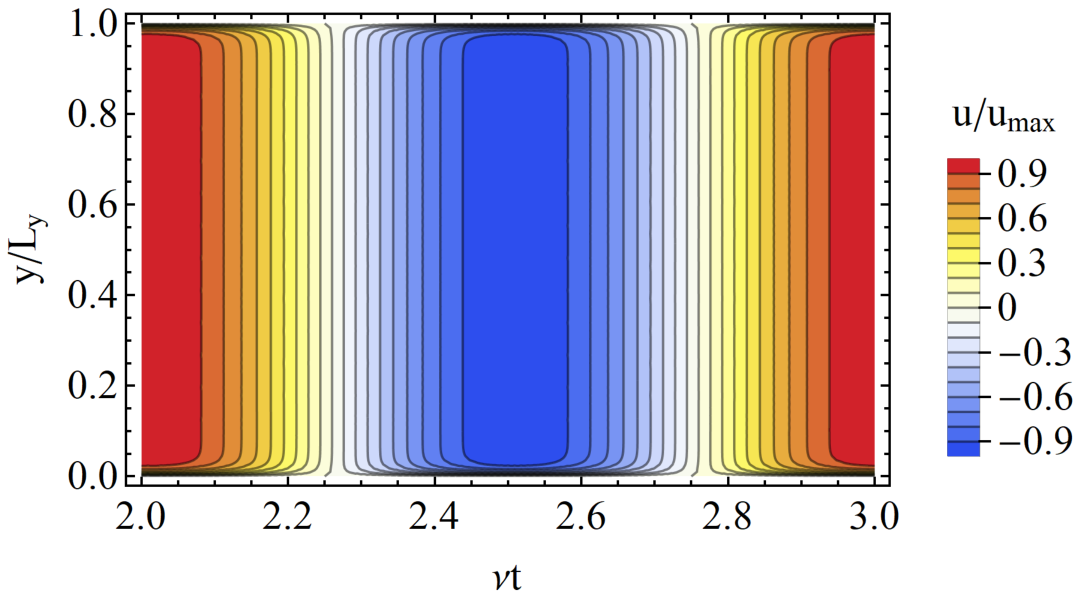}}
\hspace{0.02\textwidth}
\subfigure[]{\includegraphics[width=0.31\textwidth]{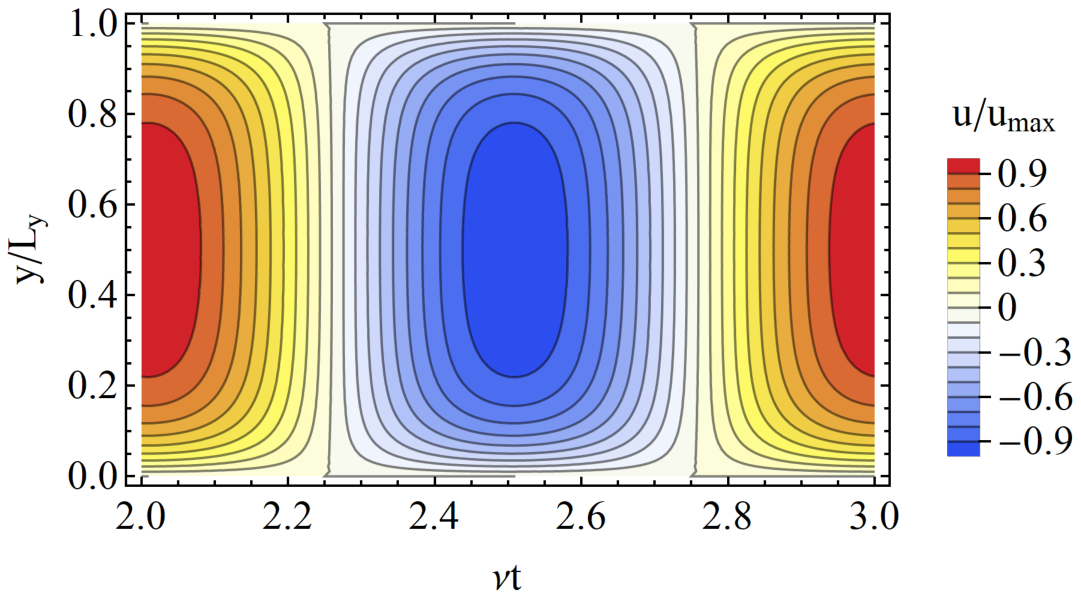}}
\hspace{0.02\textwidth}
\subfigure[]{\includegraphics[width=0.31\textwidth]{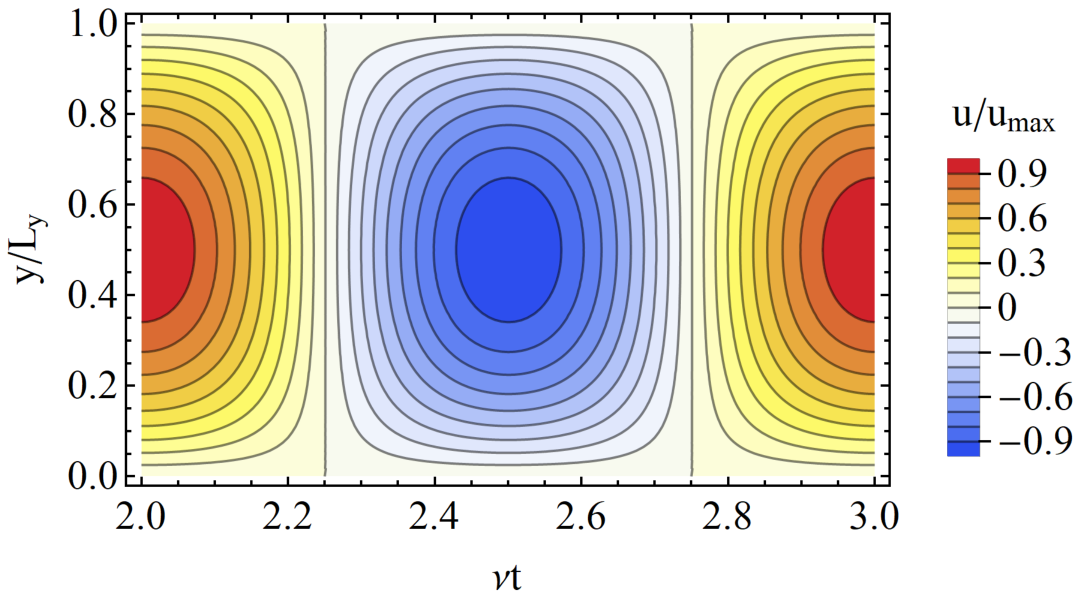}}
\caption{The normalized fluid velocity $\mathbf{u}(t,x,y)$ defined in Eq.~(\ref{dynamic-analyt-inf-ux-sol-fin}) for $\lambda_{G}/L_y=0.01$ (a), $\lambda_{G}/L_y=0.1$ (b), and $\lambda_{G}/L_y=1$ (c). In all panels, we fixed $\nu\tau=0.01$. In addition, $u_{\rm max}$ is the maximum value of the fluid velocity across the channel during the period of the oscillating driving force.
}
\label{fig:dynamic-analyt-ux-lambdaG}
\end{figure*}

In this appendix, we present the results for the normalized flow velocity at a few values of dimensionless frequency of the driving force $\nu\tau$, where $\tau$ is the momentum relaxation time, and the dimensionless Gurzhi length $\lambda_{G}/L_y$ where $L_y$ is the width of the channel; see Eq.~(\ref{steady-analyt-inf-lambdaG-def}) for the definition of the Gurzhi length. The velocity profile is defined in Eq.~(\ref{dynamic-analyt-inf-ux-sol-fin}). As one can see from Figs.~\ref{fig:dynamic-analyt-ux-nu} and \ref{fig:dynamic-analyt-ux-lambdaG}, due to the combined effect of the oscillating driving force and the no-slip boundary conditions, the velocity profiles become distorted for large $\nu\tau \gtrsim1$. If one takes a cut at a fixed value of $\nu\tau$, this would correspond to the double-peak profile in $u_x(t,y)$ shown in Fig.~\ref{fig:dynamic-analyt-u}. Another important ingredient that allows for the presence of the peaks is the viscosity quantified by the Gurzhi length. At small viscosity $\lambda_{G}/L_y \lesssim 1$, the fluid velocity profile is flat and rapidly changes only near boundaries. In this case, the peaks are hardly distinguishable, see Fig.~\ref{fig:dynamic-analyt-ux-lambdaG}(a). On the other hand, the double-peak features are suppressed for large values of $\lambda_{G}/L_y$ because the viscosity reduces the phase shift between the driving force and the fluid velocity. Thus, as we discussed in the main text [see Fig.~\ref{fig:dynamic-analyt-phase}], the double-peak profile of the hydrodynamic velocity requires large $\nu \tau$ and intermediate values of viscosity (or, equivalently, $\lambda_{G}$).

Finally, let us define the curvature of the fluid velocity profile which was used in Sec.~\ref{sec:dynamic-analyt}. We fit the velocity profile at the time when the velocity reaches its maximum via the parabolic function $a_1+a_2(y/L_y)^2$. Then, the curvature $K$ is defined as $K=-a_2/a_1$. In addition, to avoid the effects of the boundaries, we use only the part of the profile with $0.25\leq y/L_y \leq 0.75$. The curvature is maximal for the hydrodynamic profile of the current and vanishes for the Ohmic one.

\section{Role of boundary conditions}
\label{sec:app-2}

In this appendix, we investigate the role of the boundary conditions in the formation of the double-peak structure discussed in Sec.~\ref{sec:dynamic}. In general, the boundary conditions are neither no-slip nor free-surface. By using the straight infinite along the $x$ direction channel as an example, we define
\begin{eqnarray}
\label{dynamic-analyt-BC-general}
u_x(t,y=0)+ l_{\rm s, 1}\partial_y u_x(t,y=0)=0 \quad \mbox{and} \quad u_x(t,y=L_y)+ l_{\rm s, 2}\partial_y u_x(t,y=L_y)=0.
\end{eqnarray}
Here, $l_{\rm s, 1}$ and $l_{\rm s, 2}$ are the slip lengths, which could be different at different surfaces. No-slip boundary conditions correspond to $l_{\rm s}\to0$ and the free-surface ones are realized for $l_{\rm s}\to\infty$. For illustrative purposes, we consider three types of boundary conditions:
\begin{eqnarray}
\label{dynamic-analyt-BC-noslip}
\mbox{no-slip:}& \quad& u_x(t,y=0)=0, \quad u_x(t,y=L_y)=0,\\
\label{dynamic-analyt-BC-free}
\mbox{free-surface:}& \quad& \partial_y u_x(t,y=0)=0, \quad \partial_y u_x(t,y=L_y)=0,\\
\label{dynamic-analyt-BC-mixed}
\mbox{mixed:}& \quad& u_x(t,y=0)=0, \quad \partial_y u_x(t,y=L_y)=0.
\end{eqnarray}
The fluid velocity can be found along the same lines as in Sec.~\ref{sec:dynamic-analyt}. By following Ref.~\cite{Moessner-Witkowski:2018}, we use a different representation for the solution. As one can see from Eq.~(\ref{dynamic-analyt-inf-ux-sol-fin}), the time dependence of $u_x(t,y)$ is given by periodic functions and an exponential term. For $t\gg \tau$, the exponential term can be neglected. Then the Navier-Stokes equation (\ref{dynamic-analyt-inf-Eq1-x}) can be solved by replacing $\cos{(2\pi \nu t)}\to e^{- 2\pi i \nu t}$ in the driving force, solving for $u_x(t,y)= e^{-2\pi i \nu t} u_x(y)$, and taking the real part of the final result. The corresponding time-dependent solutions for the fluid velocity are
\begin{eqnarray}
\label{dynamic-analyt-ux-sol-no-slip-complex}
&&\mbox{no-slip:} \quad u_{x}(t,y) = - \frac{e \tau v_F^2 n_0 E_0}{w_0}\mbox{Re}\left\{\frac{e^{-2\pi i \nu t}}{1-2\pi i \nu \tau} \left[1 - \frac{\cosh{\left(\frac{L_y-2y}{2\lambda_{G}} \sqrt{1-2\pi i \nu \tau}\right)}}{\cosh{\left(\frac{L_y}{2\lambda_{G}} \sqrt{1-2\pi i \nu \tau}\right)}}\right] \right\},\\
\label{dynamic-analyt-ux-sol-free-complex}
&&\mbox{free-surface:} \quad u_{x}(t,y) = - \frac{e \tau v_F^2 n_0 E_0}{w_0} \mbox{Re}\left\{\frac{e^{-2\pi i \nu t}}{1-2\pi i \nu \tau}\right\},\\
\label{dynamic-analyt-ux-sol-mixed-complex}
&&\mbox{mixed:} \quad u_{x}(t,y) = - \frac{e \tau v_F^2 n_0 E_0}{w_0} \mbox{Re}\left\{\frac{e^{-2\pi i \nu t}}{1-2\pi i \nu \tau} \left[1 - \frac{\cosh{\left(\frac{L_y-y}{\lambda_{G}} \sqrt{1-2\pi i \nu \tau}\right)}}{\cosh{\left(\frac{L_y}{\lambda_{G}} \sqrt{1-2\pi i \nu \tau}\right)}}\right] \right\}.
\end{eqnarray}

\begin{figure*}[ht]
\centering
\subfigure[]{\includegraphics[width=0.31\textwidth]{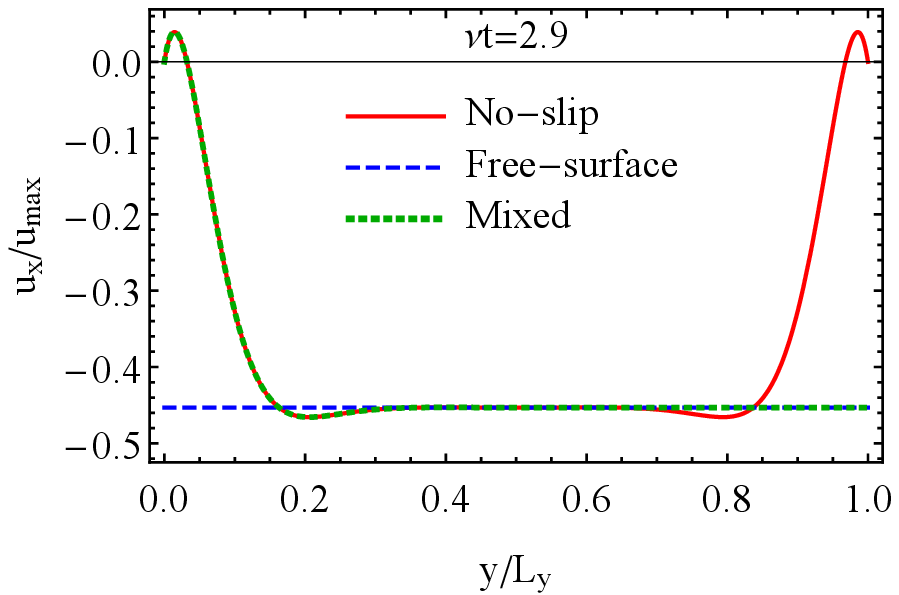}}
\hspace{0.01\textwidth}
\subfigure[]{\includegraphics[width=0.31\textwidth]{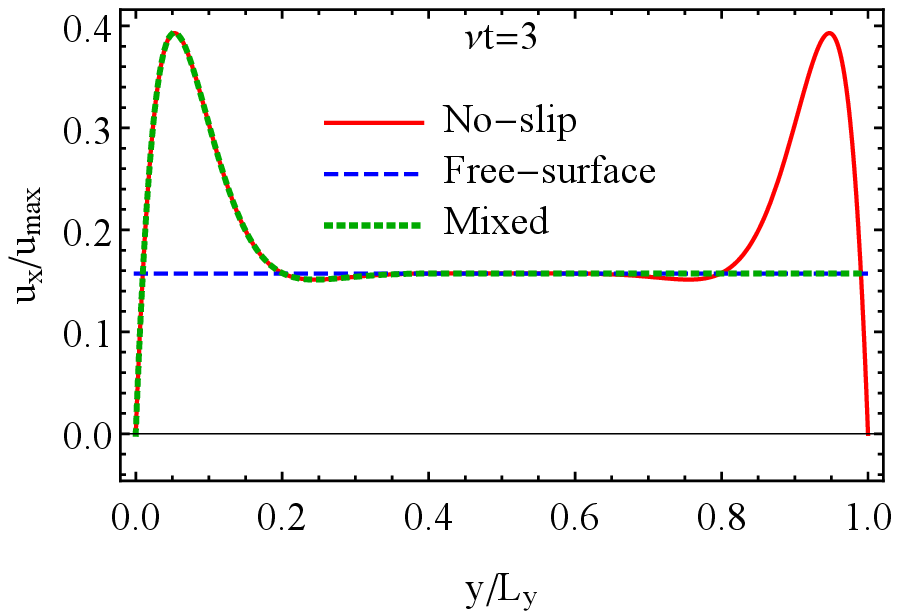}}
\hspace{0.01\textwidth}
\subfigure[]{\includegraphics[width=0.31\textwidth]{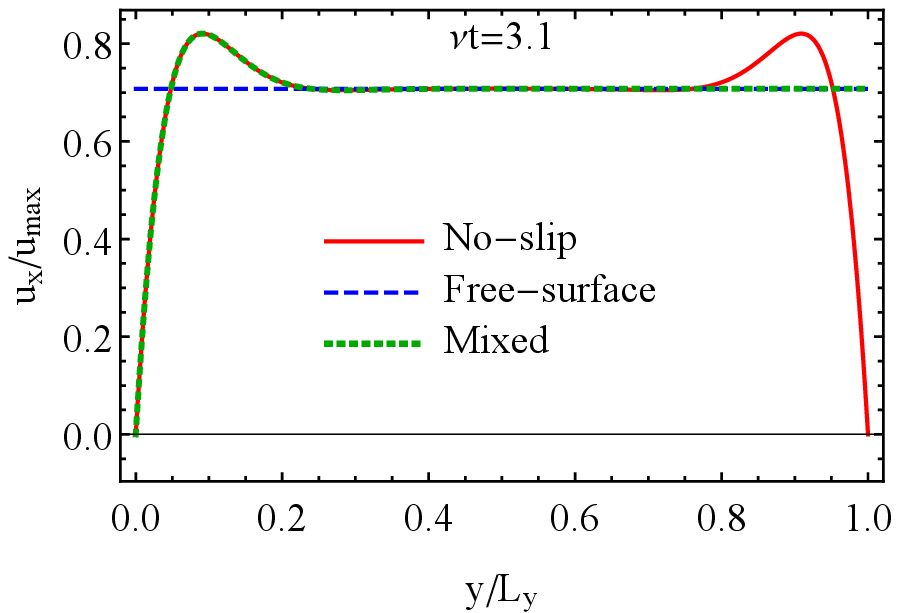}}
\caption{The normalized fluid velocity $u_x(t,y)$ for $\nu t=2.9$ (a), $\nu t=3$ (b), and $\nu t=3.1$ (c). The boundary conditions are no-slip (red solid lines), free-surface (blue dashed lines), and mixed (green thick dotted lines). In all panels, we fixed $\nu\tau=1$ and $\lambda_{G}/L_y=0.1$. We used Eqs.~(\ref{dynamic-analyt-ux-sol-no-slip-complex}), (\ref{dynamic-analyt-ux-sol-free-complex}), and (\ref{dynamic-analyt-ux-sol-mixed-complex}) for the no-slip, free-surface, and mixed boundary conditions, respectively. In addition, $u_{\rm max}$ is the maximum value of the fluid velocity across the channel during the period of the oscillating driving force.
}
\label{fig:dynamic-analyt-ux-BC-compare-all}
\end{figure*}

We compare the profile of the fluid velocities for the three boundary conditions in Fig.~\ref{fig:dynamic-analyt-ux-BC-compare-all}. As one can see, no peaks appear for the free-surface boundary conditions. In the case of the mixed boundary conditions, there is only one peak, which appears near the surface with the no-slip boundary condition. Therefore, as with the Poiseuille profile in the steady case, the peaks in the driven regime occur only when the momentum of the electron fluid is relaxed at the boundaries.

\section{Role of time profile of driving force}
\label{sec:app-3}

\begin{figure*}[ht]
\centering
\includegraphics[width=0.35\textwidth]{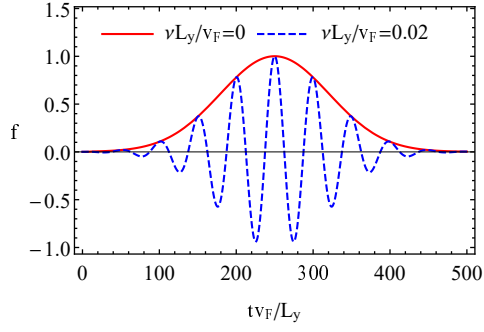}
\caption{The dependence of $f(t)$ given by Eq.~(\ref{dynamic-num-wavelet-f-def}) on time $t$ at
$\nu=0$ (red solid line) and $\nu=0.02\, v_F/L_y\approx 44~\mbox{MHz}$ (blue dashed line). We set $\xi =0.2\, v_F/L_y$, $t_{o}=10\, L_y/v_F$, $t_{\rm centr}=250\, L_y/v_F$, and $\xi_{1}=10^{-4}\, v_F^2/L_y^2$.
}
\label{fig:dynamic-num-wavelet-f}
\end{figure*}

In addition to the boundary conditions and the channel geometry, let us investigate the role of the time profile of the driving force on the formation of double peaks. For illustrative purposes, we use a wavelet profile defined as
\begin{equation}
\label{dynamic-num-wavelet-f-def}
f(t) = \frac{1}{1+e^{\xi (t_{o}-t)}} \cos{(2\pi \nu t)} e^{-\xi_1(t-t_{\rm centr})^2}.
\end{equation}
This function is plotted in Fig.~\ref{fig:dynamic-num-wavelet-f} for $\nu=0$ and $\nu=0.02\, v_F/L_y$.

\begin{figure*}[ht]
\centering
\subfigure[]{\includegraphics[height=0.225\textwidth]{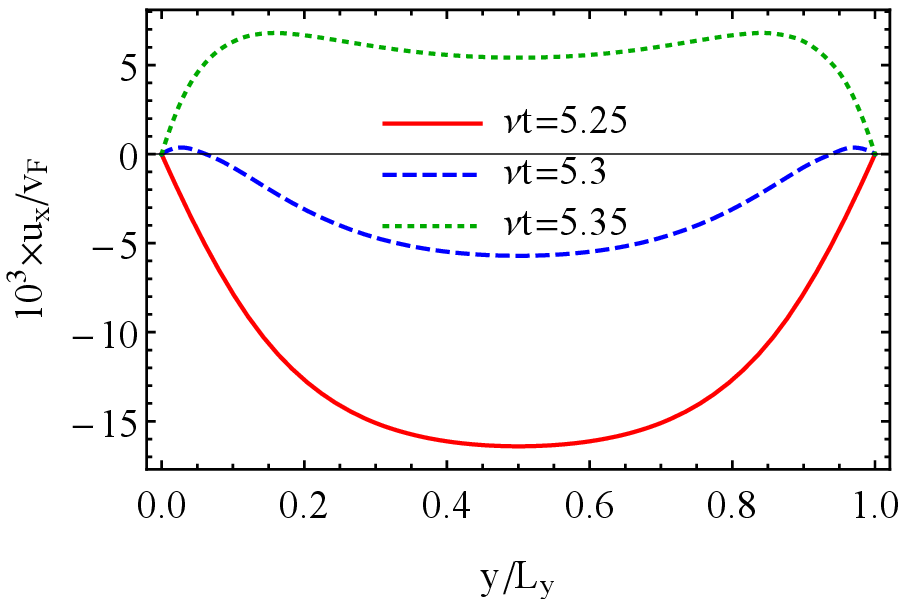}}
\hspace{0.05\textwidth}
\subfigure[]{\includegraphics[height=0.225\textwidth]{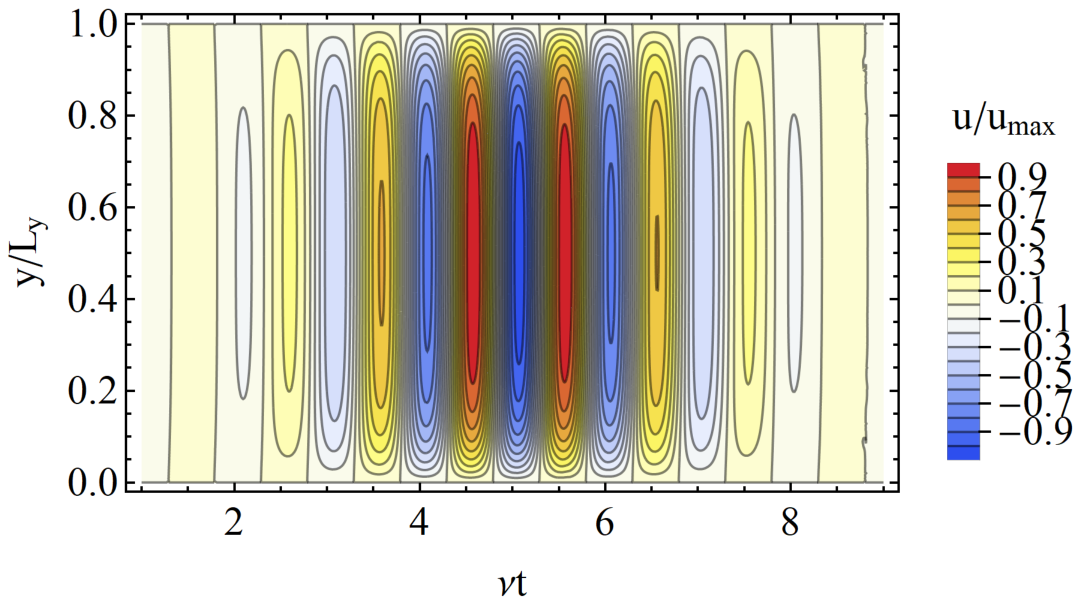}}
\caption{(a): The fluid velocity $u_x(t,x,y)$ at $x=L_x/2$ for $\nu t=5.25$ (red solid line), $\nu t=5.3$ (blue dashed line), and $\nu t=5.35$ (green dotted line). (b): The fluid velocity $u_x(t,x,y)$ at $x=L_x/2$ for several values of $t$ and $y$. In all panels, we use the channel geometry and the driving force defined in Eq.~(\ref{dynamic-num-wavelet-f-def}). In addition, $u_{\rm max}$ is the maximum value of the fluid velocity across the channel during the period of the oscillating driving force.
}
\label{fig:dynamic-num-wavelet-ux}
\end{figure*}

The numerical results for the fluid velocity in the straight channel are shown in Fig.~\ref{fig:dynamic-num-wavelet-ux}. As one can see, a double-peak profile also appears for a wavelet in the transition region. Therefore, the driving force should not be necessarily harmonic. The only requirement is the presence of transition regions where the force changes its sign.

\bibliography{library-short}

\end{document}